\documentclass[11pt]{article}
\usepackage[utf8]{inputenc}

\usepackage{amsmath,amsfonts,graphicx}
\usepackage[square,numbers,sort&compress]{natbib}
\usepackage{fullpage}
\usepackage{lineno}
\usepackage{xcolor}
\usepackage{pdflscape}
\usepackage{hyperref}
\usepackage{multicol,multirow}
\usepackage{authblk}
\usepackage{rotating}

\newcommand{\hl}{}                    

\title{How complex behavioural contagion can prevent infectious diseases from becoming endemic}
\date{}

\author[1,2]{Michael J. Plank*}
\affil[1]{School of Mathematics and Statistics, University of Canterbury, Christchurch, New Zealand}
\affil[2]{Te P\=unaha Matatini, Auckland, New Zealand}

\author[3,4,5]{Matt Ryan}
\affil[3]{School of Mathematical Sciences, Adelaide University, Adelaide, Australia}
\affil[4]{CSIRO, Adelaide and Townsville, Australia}
\affil[5]{Australian Institute of Tropical Health and Medicine, James Cook University, Townsville, Australia}

\author[6]{Lloyd Chapman}
\affil[6]{School of Mathematical Sciences, Lancaster University, Lancaster, United Kingdom}

\author[4,5]{Roslyn I. Hickson}

\author[7]{Thomas House}
\affil[7]{Department of Mathematics, University of Manchester, Manchester, United Kingdom}

\author[5]{Emma McBryde}

\author[8,9]{James M. McCaw}
\affil[8]{School of Mathematics and Statistics, The University of Melbourne, Melbourne, Australia}
\affil[9]{Centre for Epidemiology and Biostatistics, Melbourne School of Population and Global Health, The University of Melbourne, Melbourne, Australia}

\newcommand{\etal}{\textit{et al.}~}

\begin{document}
\maketitle

*Corresponding author. Email: michael.plank@canterbury.ac.nz


\begin{abstract}

Infectious disease transmission in human populations has a complex two-way interaction with changes in host behaviour. It is increasingly recognised that incorporating adaptive behavioural change into epidemic models is important for improving understanding of infectious disease dynamics and developing policy-relevant modelling tools. An important aspect of behavioural dynamics is social contagion, where people tend to adopt behaviours exhibited by others around them. In a simple behavioural contagion model, the behaviour uptake rate increases linearly with the number of contacts who have adopted a given behaviour. Here, we explore an epidemic model with complex behavioural contagion, where the behaviour uptake rate is a nonlinear function of the number of behaving contacts. We identify key bifurcation parameters of the model, which include the basic reproduction number $R_0$, the strength of the behavioural effect on disease transmission, and the speed of behaviour uptake relative to behaviour abandonment. We show that, in some regions of parameter space, the model has multiple disease-free equilibria. In this situation, the occurrence of an epidemic in a population with an initially low level of behaviour practice can trigger a self-sustaining increase in behaviour, which then causes the disease to be eliminated. In some cases, while moderate values of $R_0$ lead to the disease becoming endemic, higher values of $R_0$ may lead to behaviour-driven disease elimination. We demonstrate that this mechanism of epidemic-triggered uptake of behaviour leading to disease elimination can occur in the presence and absence of temporary post-infection immunity.

\end{abstract}

\clearpage

\section{Introduction}

As a social species, humans adjust their behaviour (e.g.,~their propensity to interact with other members of society, and how they behave when meeting them) based on their perception of how other members of the population are behaving, and their evaluation of risk (of any type) arising from that contact \cite{joffe2003risk}. In the context of infectious diseases, where risk may be defined as the probability of acquiring infection given contact, behaviour and transmission are therefore coupled. The behavioural choices of members of the population modify transmission, and in turn risk, thereby driving further changes in behavioural choices in a complex feedback loop. Behavioural changes relating to disease risk have been observed in a variety of contexts, including the Covid-19 pandemic \cite{gimma2022changes,eales2025temporal} and outbreaks of other respiratory diseases \cite{tang2004factors}, vector-borne diseases \cite{raude2019understanding} and sexually transmitted infections \cite{dewit2023sexually}.

The need to develop models of these coupled nonlinear dynamical systems, in which there is contagion of both behavioural and disease attributes, is increasingly recognised as important for understanding epidemic dynamics and developing more accurate, robust and valuable model-based policy advice \cite{marion2022modelling,hill2024integrating}.

Among a number of approaches to modelling these coupled systems (see reviews by Funk \etal \cite{funk2010modelling}, Bedson \etal \cite{bedson2021review} and Auld \etal \cite{auld2025economics}), structured compartmental models are one valuable approach. In these models, the population is compartmentalised by both disease state (e.g.,~susceptible--infectious--removed) and behavioural state (e.g.,~the `behavers' and the `non-behavers'). Changes in behavioural state may depend on the behavioural states of other individuals, as well as the observed prevalence of disease, either in the local social contact network \cite{funk2009spread} or at the population level \cite{chang2025impact}.

{\hl  There is a wide literature on social contagion in disease modelling, including models of vaccination decisions \cite{bauch2004vaccination,bauch2012evolutionary} and of behavioural change in response to perceived risk \cite{epstein2008coupled,epstein2021triple}. These studies have shown that the interaction between the dynamics of behaviour and disease transmission can qualitatively impact the epidemic dynamics \cite{kaur2025impact}. For example, social contagion of risk-avoiding behaviour can cause an epidemic to unfold in multiple waves as behaviour temporarily reverses epidemic growth when infection prevalence is sufficiently high  \cite{epstein2008coupled,perra2011towards,tyson2022transient}. In classical models, including \cite{ryan2024behaviour}, the rate of transition to a different behavioural or disease state depends linearly on the number of contacts in a given state or states. This is known as simple contagion, while complex contagion refers to a model where the dependence is nonlinear. The effects of complex social contagion in the context of infectious disease outbreaks have been studied previously, for example by modelling the spread of attitudes to vaccination \cite{campbell2013complex} or by characterising the transmission patterns arising from different contagion models \cite{contreras2024infection}.
}

Ryan \etal \cite{ryan2024behaviour} developed a model in which they explicitly drew on knowledge from the discipline of behavioural science (see Section 2.2 and Figure 2.2 of \cite{ryan2024behaviour}) to introduce a general approach to compartmental modelling for these coupled systems. Based on the socio-psychological Health Belief Model \cite{rosenstock1966people, becker1974health}, they derived a functional form for behaviour uptake that depends on sociality and perception-informed infection feedback. They applied this to an epidemic transmission model to study the detailed dynamics of a particular model.

Here, we draw on that socio-psychological belief formalism and, through an alternative line of reasoning, suggest an additional plausible case of their general framework, and study the resultant model's dynamics.
In particular, while Ryan \etal \cite{ryan2024behaviour} assumed the rate of behaviour uptake depended linearly on both the prevalence of infection and the prevalence of behavers, we propose that this second term may be nonlinear. This is a type of complex social contagion, reflecting an additional level of self-reinforcement of visible behaviour in which behaviour transmission depends on interactions with multiple behavers \cite{centola2007complex}. This is well motivated since social complex contagion spread has been observed in real world behavioural-epidemic systems \cite{centola2010spread, sprague2017evidence}.
We characterise the structural properties of two examples of this generalised behaviour--disease model, one in which those who `behave' have a reduced susceptibility, and one in which they have a reduced contagiousness.

We focus on simple compartmental models that are amenable to mathematical analysis. Our aim is not to model any specific disease or outbreak situation, but rather to characterise the qualitative dynamics and explore phenomena that do not occur in comparable models without a behavioural component, or in models such as \cite{ryan2024behaviour,ryan2026behaviour} where behaviour follows a simple contagion process. {\hl The novel contribution we make is not the use of complex contagion to represent behaviour, but its consequences for the equilibrium structure of a coupled behaviour--disease system.} We show that complex contagion of behaviour results in a system that admits multiple disease-free equilibria and dynamical regimes that cannot arise under simple behavioural contagion. This includes situations where an initial rise in infections triggers a self-sustaining increase in behaviour that is sufficient to eliminate infection. We show that this mechanism can occur whether or not recovery from infection induces temporary immunity.

\section{Methods}

We consider a modified version of the `behaviour and disease model' proposed in \cite{ryan2024behaviour}, with two main modifications: we focus primarily on a model in which infected individuals immediately become susceptible following the end of the infectious period (although later we also consider a model in which recovery from infection confers temporary immunity); and, importantly, we assume that the social behaviour uptake rate is proportional to $B^2$ rather than $B$, where $B$ is the total fraction of the population in the behaving class. This is a susceptible-infectious-susceptible (SIS) model with behavioural dynamics that are a type of complex social contagion \cite{centola2007complex}, meaning that the behaviour uptake rate depends nonlinearly (in this case quadratically) on the number of contacts who are in the behaving class.

\subsection{SIS model}

{\hl We assume that, at a given point in time, the population can be compartmentalised into those who are practising a particular behaviour (whom we refer to as ``behavers'') and those who are not.} We assume that the susceptibility of behavers is reduced by a factor $q_s\in[0,1]$ and their contagiousness is reduced by a factor $q_c\in[0,1]$, relative to non-behavers. For convenience, we work with dimensionless time $t$ measured in units of the average infectious period (i.e.,~the rate at which infectious individuals become susceptible is unity). The model is described by a system of four ordinary differential equations, for the proportion of the population in each of the four compartments: susceptible non-behavers ($S_N$), infectious non-behavers ($I_N$), susceptible behavers ($S_B$), and infectious behavers ($I_B$)
\begin{align} 
\frac{dS_N}{dt} &= -\lambda S_N +  I_N - \omega(S,B) S_N + \alpha S_B \label{eq:model1} \\
\frac{dI_N}{dt} &= \lambda S_N -  I_N - \omega(S,B) I_N + \alpha I_B \\
\frac{dS_B}{dt} &= - (1-q_s) \lambda S_B +  I_B + \omega(S,B) S_N - \alpha S_B \\
\frac{dI_B}{dt} &=  (1-q_s) \lambda S_B -  I_B + \omega(S,B) I_N - \alpha I_B \label{eq:model4}
\end{align}
where the force of infection is 
\begin{equation} \label{eq:FOI}
\lambda = R_0 \left(I_N + (1-q_c)I_B\right)
\end{equation}
{\hl Note that for simplicity, we have parametrised the model in terms of the basic reproduction number $R_0$ as opposed to a transmission rate parameter, commonly denoted $\beta$.} We define the rate of behaviour uptake per unit of dimensionless time as 
\begin{equation} \label{eq:omega}
\omega(S,B) = \alpha\left( \tau B^2 + \chi (1-S) \right)
\end{equation}
where $S = S_N + S_B$ is the susceptible fraction and $B = S_B + I_B$ is the behaving fraction. This definition of $\omega$ contains a social contagion term $\tau B^2$ and a disease-induced term $\chi I = \chi(1-S)$, while $\alpha$ is a rate constant representing the speed of behavioural change. Note that $\alpha$ appears as a common factor in the rates of behaviour abandonment, social behaviour uptake, and disease-induced terms. This simplifies the subsequent analysis, but is not a restrictive assumption since the parameters $\tau$ and $\chi$ allow all three rates to be set independently.

The assumption of quadratic dependence of the social contagion term on $B$ models a process where social behaviour uptake requires a non-behaver to interact with two behaving contacts, as opposed to one behaving contact for a simple contagion process. This can be thought of as a mass action process $N+2B \rightarrow B+2B$, in contrast to $N+B \rightarrow B+B$ for a simple contagion. {\hl There is some evidence for this qualitative type of contagion in the behavioural literature in certain contexts \cite{centola2010spread,sprague2017evidence}}. However, other models of complex contagion are possible, such as uptake rates that have a threshold or saturating relationship with respect to $B$. {\hl Another alternative would be to include one or more latent behavioural compartments, in which individuals have become predisposed to adopting the behaviour but need further interaction with behavers to do so. However, this would add substantial model complexity and we do not pursue this further. }

Since the total population size is conserved, Eqs. \eqref{eq:model1}--\eqref{eq:model4} can be reduced to a three-dimensional system. It is convenient to express this as a system of differential equations for $S_N$, $S_B$ and $B=S_B+I_B$ (retaining use of $S = S_N + S_B$ for notational convenience):
\begin{align}
\frac{dS_N}{dt} &= -\lambda S_N +  1-B-S_N - \omega S_N + \alpha S_B \label{eq:3D_system1} \\
\frac{dS_B}{dt} &= -(1-q_s)\lambda S_B +  B-S_B + \omega S_N - \alpha S_B \label{eq:3D_system2} \\
\frac{dB}{dt} &= \omega(S,B) (1-B) - \alpha B, \label{eq:3D_system3}
\end{align}
where $\omega$ is defined by Eq. \eqref{eq:omega} and the force of infection $\lambda$ is given by
\begin{equation} \label{eq:FOI2}
\lambda = R_0 \left(1-S - q_c (B-S_B)\right)
\end{equation} 

We consider two special cases of this model, one where the behaviour solely reduces {\hl contagiousness} (i.e.,~$q_s=0$) and one where it solely reduces susceptibility (i.e.,~$q_c=0$). We refer to these as the {\hl contagiousness-modulated} and susceptibility-modulated models respectively. {\hl We will show that the contagiousness-modulated model admits a two-dimensional reduction, which we will explore using a standard bifurcation analysis. We will calculate numerical solutions of Eqs. \eqref{eq:3D_system1}--\eqref{eq:FOI2} to verify our analytical findings and to extend our analysis to the susceptibility-modulated model. } 

To plot phase portraits of the reduced two-dimensional model, we compute the equilibria and associated Jacobian matrix numerically. In cases where the system is bistable, the boundary separating the basins of attraction of the stable equilibria is the stable manifold of a saddle equilibrium. We calculate this by perturbing from the saddle equilibrium in the stable eigendirection and solving the differential equations backwards in time. Similarly, we compute the unstable manifold by perturbing in the unstable eigendirection and solving the equations forwards in time.

\subsection{SIRS model}
Although we will focus primarily on the SIS model in which individuals immediately become susceptible at the end of their infectious period, we also consider a version of the model where recovery from infection confers temporary immunity. We model this in the standard way by introducing a recovered compartment, and assuming that individuals transition from recovered to susceptible at constant rate $\nu>0$, independently of whether they are behaving or not. This implies that the average immune period, measured in units of the infectious period, is $1/\nu$. As before, we assume that behavioural transitions are independent of disease status (see Supplementary Material sec. 1 for details).

\section{Results}

\subsection{Contagiousness-modulated SIS model: analytical results} \label{sec:analytical_results}

In the contagiousness-modulated model (i.e., when $q_s=0$), behavioural transitions are independent of disease state, and disease state transitions are independent of behavioural state. {\hl As a consequence, the dynamics of Eqs. \eqref{eq:3D_system1}--\eqref{eq:3D_system3} relax onto the invariant two-dimensional subsystem in which $S_B(t) = S(t)B(t)$ (see Supplementary Material sec 2.1 for details). The equilibria and stability of the full model must therefore be the same as those of the two-dimensional system that results from substituting $S_B(t) = S(t)B(t)$ into Eqs. \eqref{eq:3D_system1}--\eqref{eq:3D_system3} and writing the system in terms of $S(t)$ and $B(t)$:}
\begin{align}
\frac{dS}{dt} &= -R_0 S(1-S)\left(1-q_c B\right) +  (1-S) \label{eq:2Dsystem1} \\
\frac{dB}{dt} &=  \alpha \left[ \left(\tau B^2 + \chi (1-S)\right) (1-B) -  B\right]
\label{eq:2Dsystem2}
\end{align}
{\hl In Section \ref{sec:numerical_results}, we will verify that the full model has the same equilibrium solutions by solving Eqs. \eqref{eq:3D_system1}--\eqref{eq:3D_system3} numerically for a range of parameter values and initial conditions.}

Eqs. \eqref{eq:2Dsystem1}--\eqref{eq:2Dsystem2} have a disease-free equilibrium (DFE) with no behaviour at $(S^*,B^*)=(1,0)$ and, as for classical models, the disease can invade from this equilibrium if and only if $R_0 > 1$. We refer to this equilibrium as the no-behaviour DFE (NDFE). 
Any other disease-free equilibria must have $S^*=1$ and, from Eq. \eqref{eq:2Dsystem2}, must satisfy $\tau B^*(1-B^*)=1$, which has roots at
\begin{equation} \label{eq:BDFE}
B^*_\pm = \frac{1}{2}\left(1 \pm \sqrt{1-4/\tau}\right)
\end{equation}
We refer to these two roots as the DFEs with behaviour and denote them BDFE$_\pm$. The dimensionless constant $\tau$, which represents the rate of social behaviour uptake relative to the rate of behaviour abandonment, is a key bifurcation parameter in the model. {\hl The role of $\tau$ is clearest if $\tau B (1-B)$ is read as the behavioural counterpart of the effective reproduction number: the average number of replacements that a behaver recruits before abandoning the behaviour, with the factor $B$ the chance that the second contact required for recruitment is also a behaver and the factor $(1-B)$ the chance that the recruit is not. Because two behaving contacts are needed, this replacement number vanishes when behaviour is rare, rather than peaking as it would under simple contagion, so behaviour can sustain itself only when the replacement number exceeds $1$.} When $\tau<4$ (i.e.,~social behaviour uptake is low relative to abandonment), the only disease-free equilibrium is the NDFE with $B^*=0$. As $\tau$ increases through $\tau=4$, two additional disease-free equilibria with $B^*>0$ are created via a saddle node bifurcation at $(S^*,B^*)=(1,1/2)$. 

In the invariant disease-free subsystem (i.e.,~when $S=1$), the NDFE is always stable to perturbations in $B$ and the equilibria BDFE$_\pm$ (Eq. \eqref{eq:BDFE}) are stable and unstable respectively to perturbations in $B$ {\hl (see Supplementary Figure S1)}. The stability of these equilibria with respect to perturbation in the $S$ direction (i.e.,~to introduction of disease) depends on model parameters as now described.

At each of the two BDFEs, the reproduction number is
\begin{equation} \label{eq:R_BDFE}
R^*_\pm = R_0 \left(1 -q_c B^*_\pm \right)
\end{equation}
The disease can invade BDFE$_\pm$ if and only if $R^*_\pm>1$, which is equivalent to the condition $R_0>R_{0,TC\pm}$ where 
\begin{equation} \label{eq:R0TC}
    R_{0,TC\pm} = \frac{1}{1 -q_c B^*_\pm}
\end{equation}
The values of $R_{0,TC\pm}$ define two transcritical bifurcations which occur at BDFE$_\pm$. In each transcritical bifurcation, the relevant BDFE loses stability and an endemic equilibrium (EE) moves into the biological region of phase space, i.e.,~$(S,B)\in[0,1]^2$.

{\hl Setting $q_c=1$, so that behavers do not transmit at all, gives $R^*_+=R_0(1-B^*_+)=R_0B^*_-$, since the two roots of Eq. \eqref{eq:BDFE} sum to $1$. Even a perfectly protective behaviour therefore leaves BDFE$_+$ stable only for $R_0<1/(1-B^*_+)$, which is the equivalent of the classical herd immunity condition $R_0<1/S$, but with the non-behaving fraction in place of the susceptible fraction.}

Saddle node bifurcations (SNBs) of EEs occur when the $S$ and $B$ nullclines intersect tangentially (see Supplementary Material sec. 2.2 and Figure \ref{fig:nullclines}). This can happen on the upper or lower branch of the sigmoidally-shaped $B$ nullcline, which we denote SNB1 and SNB2 respectively. SNB1 only exists if $\tau<4$ (i.e.,~when the BDFEs do not exist). 

\begin{figure}
    \centering
    \includegraphics[trim={2cm 0 0cm 0},clip,width=\linewidth]{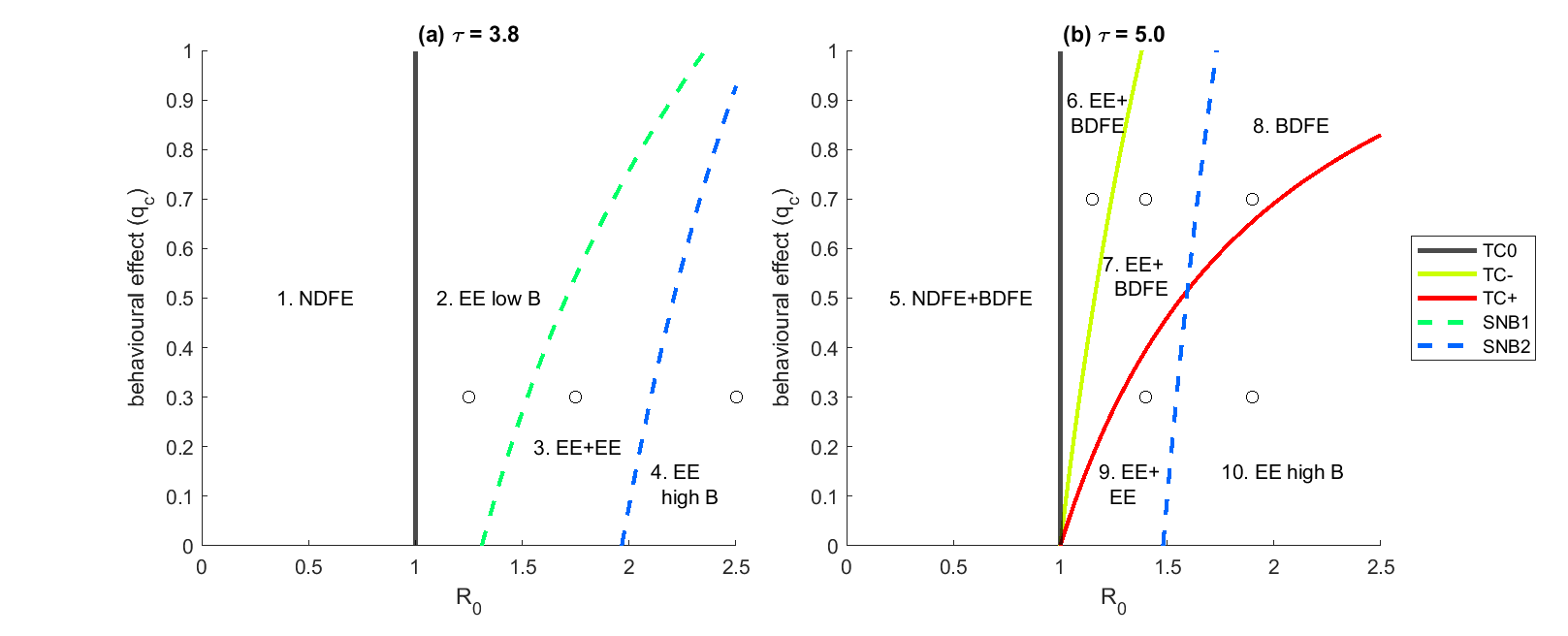}
    \caption{Two-parameter bifurcation diagram $(R_0,q_c)$ for the contagiousness-modulated model: (a) when $\tau<4$ (i.e.,~no BDFEs exist); (b) when $\tau>4$ (i.e.,~two BDFEs exist). $\chi=0.2$ in both plots.  The black vertical line is the standard disease invasion threshold at $R_0=1$, which is a transcritical bifurcation (TC0) of the NDFE and an EE. Solid yellow and red curves are transcritical bifurcations (TC$-$ and TC$+$ respectively) of a BDFE and EE. Dashed green and blue curves are saddle node bifurcations (SNB1 and SNB2 respectively) of two EEs. Each region is labelled by its regime number from Table \ref{tab:regimes} and its stable equilibria; where two equilibria are listed within a region, the dynamics are bistable.  An epidemic followed by disease dying out due to behaviour can occur in (b) in the regions labelled `EE+BDFE' and `BDFE', but is only guaranteed to occur in the latter region (whereas in the `EE+BDFE' region it only occurs if there is enough behaviour at the initial condition). Open black circles show the parameter combinations for which phase plots and time series solutions are shown in Figures \ref{fig:cases1}--\ref{fig:cases2}. Supplementary Figure S3 shows one-parameter ($R_0$) bifurcation diagrams corresponding to a horizontal cross-section through these plots. }
    \label{fig:bifurcation}
\end{figure}

The three key bifurcation parameters for the contagiousness-modulated SIS model \eqref{eq:2Dsystem1}--\eqref{eq:2Dsystem2} are the basic reproduction number $R_0$, the strength of the behavioural effect $q_c$, and the rate of social behaviour uptake relative to behaviour abandonment $\tau$. Changes in the values of these parameters can change the number of equilibria of the model and/or their stability. The model has ten distinct parameter regions (Table \ref{tab:regimes}), separated by the bifurcations described above and shown in Figure \ref{fig:bifurcation}. Supplementary Material sec. 2.3 provides further details and an interactive bifurcation diagram is available as Electronic Supplementary Material and at \cite{github-repo}. {\hl In Section \ref{sec:numerical_results}, we explain how the qualitative dynamics of the model change in the different parameter regimes shown in Figure \ref{fig:bifurcation} and verify these findings using numerical solutions of Eqs. \eqref{eq:3D_system1}--\eqref{eq:3D_system3}. }

\begin{landscape}
\begin{table}[]
{\scriptsize
    \centering
    \begin{tabular}{lllllllllp{5.5cm}}
    \hline
     & $R_0$ & $\tau$ &  NDFE & BDFE$_-$ & BDFE$_+$ & EE low & EE med & EE high & Dynamics \\
    \hline
1. & $R_0<1$ & $\tau < 4$ &  \underline{Stable} & - & - & - & - & - & Neither disease nor behaviour can invade \\
2. &  $1<R_0<R_{0,SNB1}$ &  $\tau < 4$ & Saddle & - & - &  \underline{Stable} & - & - &  Disease invades and becomes endemic (low $B$) \\
3. &  $R_{0,SNB1}<R_0<R_{0,SNB2}$ &  $\tau < 4$ & Saddle & - & - &  \underline{Stable} & Saddle &  \underline{Stable} &  Disease invades and becomes endemic (bistable low/high $B$) \\
4. &  $R_{0,SNB2}<R_0$ &  $\tau < 4$ & Saddle & - & - & - & - &  \underline{Stable} &  Disease invades and becomes endemic (high $B$) \\
\hline
5. & $R_0<1$ &  $\tau > 4$  &  \underline{Stable} & Saddle &  \underline{Stable} & - & - & - & Disease cannot invade (bistable zero or high $B$) \\
6. & $1<R_0<R_{0,TC-}$ & $\tau > 4$  & Saddle & Saddle &  \underline{Stable} & \underline{Stable} & - & - & Disease either invades and becomes endemic or dies out (with no outbreak) due to behaviour persisting (bistable).  \\
7. & $R_{0,TC-}<R_0<\left\{R_{0,SNB2},R_{0,TC+}\right\}$ &  $\tau > 4$  & Saddle & Unstable &   \underline{Stable} &  \underline{Stable} & Saddle & - & Disease invades and can either become endemic or die out due to behaviour persisting (bistable)\\
8. & $R_{0,SNB2}<R_0<R_{0,TC+}$& $\tau > 4$  & Saddle & Unstable & \underline{Stable} &  - & - & - & Disease invades but then dies out due to behaviour persisting\\
9. & $R_{0,TC+}<R_0<R_{0,SNB2}$ &  $\tau > 4$  & Saddle & Unstable & Saddle &   \underline{Stable} & Saddle &  \underline{Stable} & Disease invades and becomes endemic (bistable low/high $B$) \\
10. & $\left\{R_{0,SNB2},R_{0,TC+}\right\}<R_0$  & $\tau > 4$ & Saddle & Unstable & Saddle &  - & - &  \underline{Stable} & Disease invades and becomes endemic (high $B$) \\
\hline
    \end{tabular}
    \caption{Classification of parameter regimes of the contagiousness-modulated model according to the existence and stability of different equilibria, depending on the values of the basic reproduction number $R_0$ and the rate of social behaviour uptake relative to behaviour abandonment  $\tau$. Stable equilibria are indicated by underlining; `-' indicates that the equilibrium does not exist in the biological region of phase space (i.e.,~$(S,B)\in[0,1]^2$). Bistable dynamics occur when two different equilibria are stable. The NDFE and BDFE$_+$ are always stable to perturbations in behaviour and, when they are of saddle-type, are unstable to introduction of disease (i.e.,~the stable manifold is the $B$ axis and unstable manifold is the $S$ axis). BDFE$_-$ is always unstable to perturbations in behaviour and, when it is of saddle type, is stable to introduction of disease. When `EE med' is of saddle-type, its stable and unstable manifolds are nonlinear invariant curves (see phase plane graphs in Figures \ref{fig:cases1}--\ref{fig:cases2}).  $R_{0,TC\pm}$ are the two values of $R_0$ in Eq. \eqref{eq:R0TC} at which there is a transcritical bifurcation of BDFE$_{\pm}$ with an endemic equilibrium. $R_{0,SNB1}$ and $R_{0,SNB2}$ are the values of $R_0$ at which there are saddle node bifurcations of endemic equilibria (SNB1 and SNB2). Note these bifurcation points are functions of the parameters $q_c$, $\tau$ and $\chi$ (Figure \ref{fig:bifurcation}).}
    \label{tab:regimes}
    }
\end{table}
\end{landscape}

\subsection{Contagiousness-modulated SIS model: numerical results}
\label{sec:numerical_results}

\begin{figure}
    \centering
    \includegraphics[trim={0.5cm 1.5cm 1cm 0.3cm},clip,width=\linewidth]{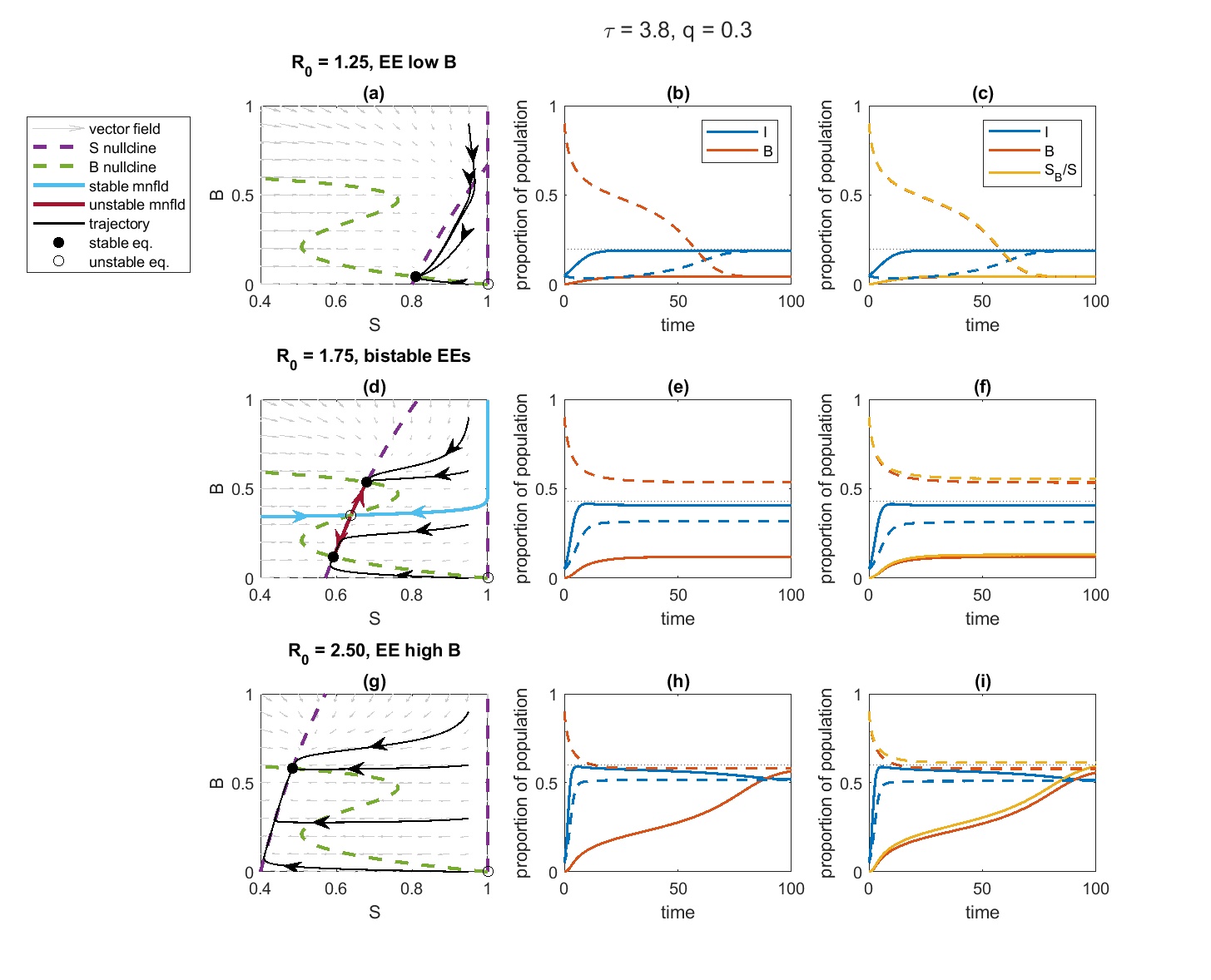}
    \caption{\scriptsize Dynamics in parameter regimes 2 (a,b,c), 3 (d,e,f) and 4 (g,h,i) for: the contagiousness-modulated model with $q_c=0.3$ shown as phase plots in the $(S,B)$ plane (left column) and time series solutions (centre column); and for the susceptibility-modulated model with $q_s=0.3$ shown as time series solutions (right column). All plots have $\tau=3.8$ (so there are no BDFEs), $\chi=0.2$, and $\alpha=0.25$.  Phase plots show the vector field (grey arrows), $S$ nullcline (dashed purple), $B$ nullcline (dashed green), the stable manifold (solid blue) and unstable manifold (solid red) of the saddle endemic equilibrium (when it exists), stable equilibria (filled circles), unstable equilibria (open circles), and trajectories for selected initial conditions (solid black). Time series plots (centre and right columns) show the prevalence of infection (blue) and of behavers (red) over time, for two initial conditions, one with $B(0)=0$ (solid curves) and one with $B(0)=0.9$ (dashed curves). Both initial conditions have $S(0)=0.95$ and $S_B(0)=S(0)B(0)$. Plots for the susceptibility-modulated model (right column) also show the proportion of the susceptible population that is behaving (yellow). In bistable cases, the two initial conditions converge to two different stable equilibria. Otherwise they converge to the unique stable equilibrium. The dotted horizontal line shows the equilibrium prevalence ($I=1-1/R_0$) for the case where there is no behaviour in the model.}
    \label{fig:cases1}
\end{figure}

When the social behaviour uptake rate is low ($\tau<4$), the only disease-free equilibrium is that with $B=0$ (the NDFE). If $R_0<1$, the disease cannot invade and the NDFE is globally stable. If $R_0>1$, the NDFE is unstable and the system tends towards a stable endemic equilibrium. If $R_0$ is only slightly above $1$ or the behavioural effect $q_c$ is sufficiently strong, this equilibrium has {\hl low prevalence of behavers} (Figure \ref{fig:cases1}a--b). If $R_0$ is large or $q_c$ is small, it has {\hl high prevalence of behavers} (Figure \ref{fig:cases1}g--h). For intermediate values of $R_0$ and $q_c$, both high- and low-behaviour endemic equilibria are stable and the long-term outcome depends on initial conditions (Figure \ref{fig:cases1}d--e).  

\begin{figure}
    \centering
    \includegraphics[trim={0.5cm 1.5cm 1cm 0.3cm},clip,width=\linewidth]{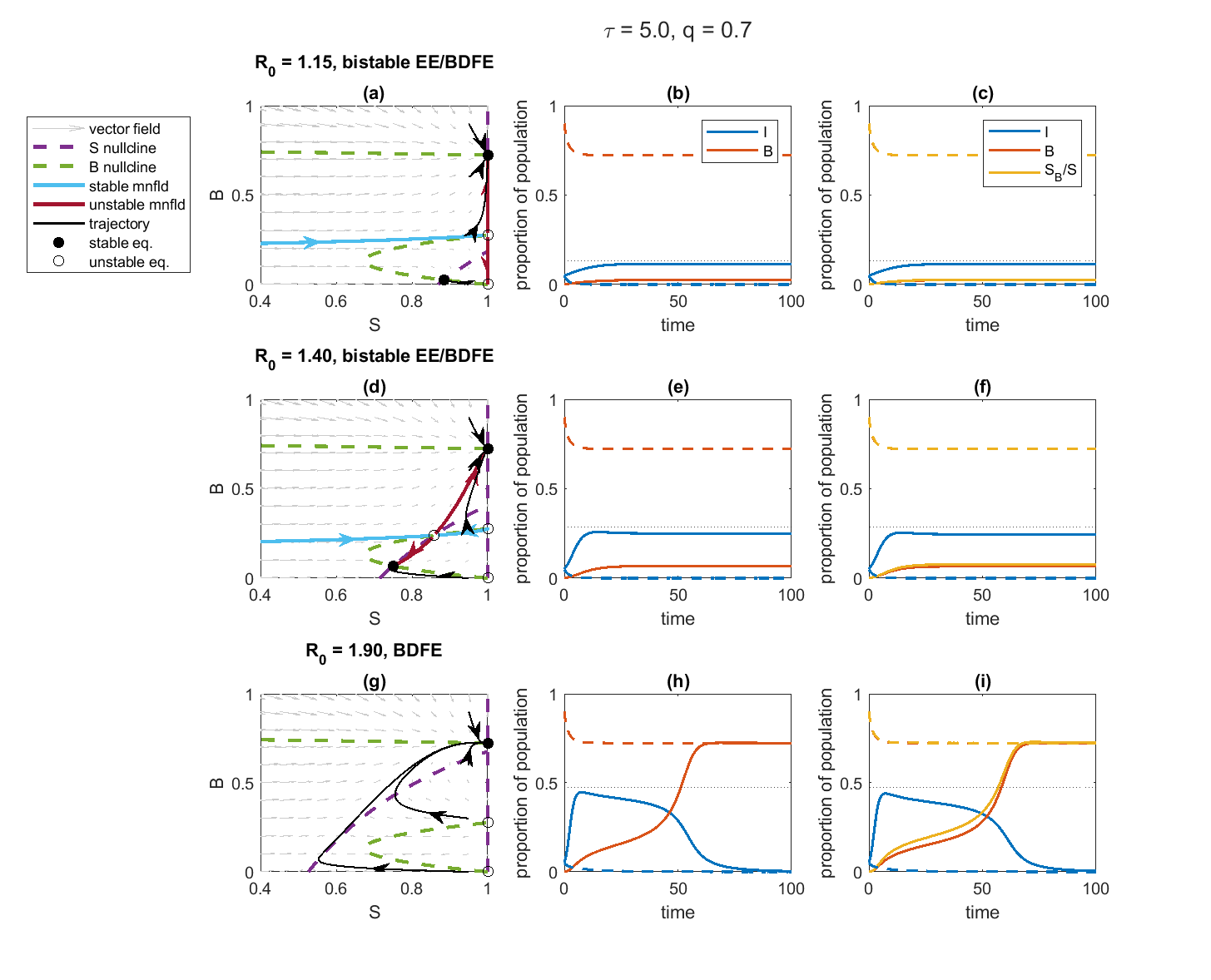}
    \caption{\scriptsize Dynamics in parameter regimes 6 (a,b,c), 7 (d,e,f) and 8 (g,h,i) for: the contagiousness-modulated model with $q_c=0.7$ shown as phase plots in the $(S,B)$ plane (left column) and time series solutions (centre column); and for the susceptibility-modulated model with $q_s=0.7$ shown as time series solutions (right column). All plots have $\tau=5$ (so there are two BDFEs), $\chi=0.2$, and $\alpha=0.25$.  Phase plots show the vector field (grey arrows), $S$ nullcline (dashed purple), $B$ nullcline (dashed green), the stable manifold (solid blue) and unstable manifold (solid red) of the saddle endemic equilibrium (when it exists), stable equilibria (filled circles), unstable equilibria (open circles), and trajectories for selected initial conditions (solid black). Time series plots (centre and right columns) show the prevalence of infection (blue) and of behavers (red) over time, for two initial conditions, one with $B(0)=0$ (solid curves) and one with $B(0)=0.9$ (dashed curves). Both initial conditions have $S(0)=0.95$ and $S_B(0)=S(0)B(0)$. Plots for the susceptibility-modulated model (right column) also show the proportion of the susceptible population that is behaving (yellow). In bistable cases, the two initial conditions converge to two different stable equilibria. Otherwise they converge to the unique stable equilibrium. The dotted horizontal line shows the equilibrium prevalence ($I=1-1/R_0$) for the case where there is no behaviour in the model.}
    \label{fig:cases3}
\end{figure}

\begin{figure}
    \centering
    \includegraphics[trim={0.5cm 0.5cm 1cm 0.3cm},clip,width=\linewidth]{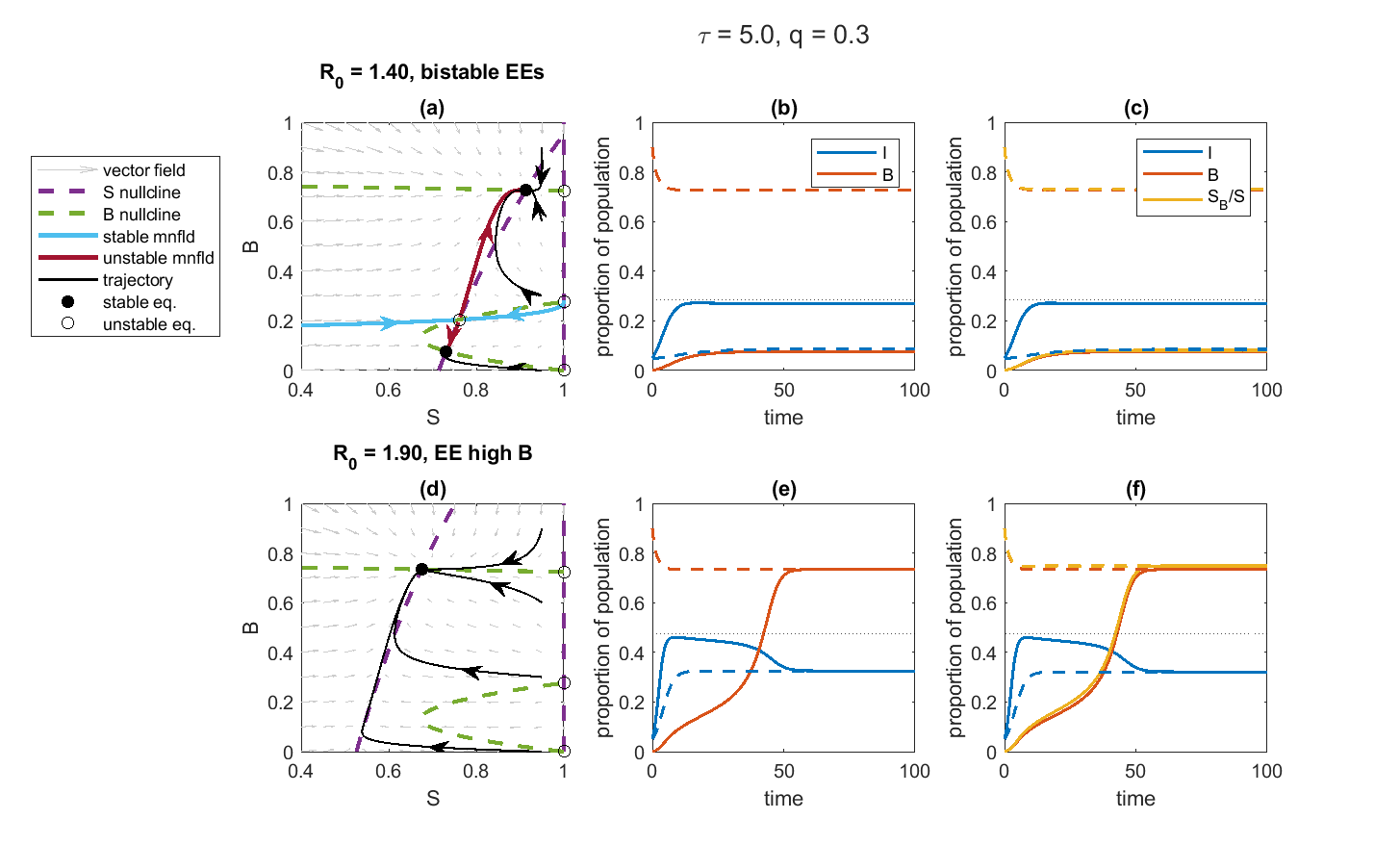}
    \caption{\scriptsize Dynamics in parameter regimes 9 (a,b,c) and 10 (d,e,f) for: the contagiousness-modulated model with $q_c=0.3$ shown as phase plots in the $(S,B)$ plane (left column) and time series solutions (centre column); and for the susceptibility-modulated model with $q_s=0.3$ shown as time series solutions (right column). All plots have $\tau=5$ (so there are two BDFEs), $\chi=0.2$, and $\alpha=0.25$.  Phase plots show the vector field (grey arrows), $S$ nullcline (dashed purple), $B$ nullcline (dashed green), the stable manifold (solid blue) and unstable manifold (solid red) of the saddle endemic equilibrium (when it exists), stable equilibria (filled circles), unstable equilibria (open circles), and trajectories for selected initial conditions (solid black). Time series plots (centre and right columns) show the prevalence of infection (blue) and of behavers (red) over time, for two initial conditions, one with $B(0)=0$ (solid curves) and one with $B(0)=0.9$ (dashed curves). Both initial conditions have $S(0)=0.95$ and $S_B(0)=S(0)B(0)$. Plots for the susceptibility-modulated model (right column) also show the proportion of the susceptible population that is behaving (yellow). In bistable cases, the two initial conditions converge to two different stable equilibria. Otherwise they converge to the unique stable equilibrium. The dotted horizontal line shows the equilibrium prevalence ($I=1-1/R_0$) for the case where there is no behaviour in the model.}
    \label{fig:cases2}
\end{figure}

When the social behaviour uptake rate is high ($\tau>4$), there are three disease-free equilibria: one with $B=0$ and two with $B>0$. As before, if $R_0<1$ the disease cannot invade. If $R_0$ is slightly above $1$ or the behavioural effect $q_c$ is sufficiently strong, the DFE with high {\hl prevalence of behavers} is stable. Depending on the exact position in the $(R_0,q_c)$ parameter space (Figure \ref{fig:bifurcation}b), there may also be a stable endemic equilibrium, in which case the disease may either die out or become endemic in the long-term depending on initial conditions (Figure \ref{fig:cases3}a,b,d,e). Alternatively, the DFE with high {\hl prevalence of behavers} may be the only stable equilibrium, in which case the disease always dies out (Figure \ref{fig:cases3}g,h). If $R_0$ is sufficiently large or {\hl the behavioural effect} $q_c$ is sufficiently weak, all the DFEs are unstable and the disease always becomes endemic (Figure \ref{fig:cases2}). 

Note that, even in the situations that result in endemicity of disease, the equilibrium infection prevalence $I^*=1-S^*$ is necessarily less than in a classical SIS model without behaviour included (i.e.,~$I^*$ is always less than $1-1/R_0$, indicated by the horizontal dotted line in the middle column of plots in Figures \ref{fig:cases1}--\ref{fig:cases2}). If disease is present, then behaviour is as well, and therefore prevalence is lower than would otherwise be the case. {\hl Supplementary Figure S3 shows the infection prevalence as a function of $R_0$ for the different equilibria in the $\tau<4$ and $\tau>4$ cases.}

Interestingly, this analysis reveals that there are situations where a higher value of $R_0$ can lead to disease extinction while a lower value may result in endemicity, and thus perhaps a greater long-term burden of disease. {\hl This occurs because sufficiently high infection prevalence is needed to increase the number of behavers above a critical threshold, which subsequently causes the disease to die out.} This can be seen by comparing parameter regime 7 (see Figure \ref{fig:cases3}d,e where $R_0=1.4$ and initial conditions with low $B$ result in endemicity) with parameter regime 8 (see Figure \ref{fig:cases3}g,h where $R_0=1.8$ and the disease always dies out eventually) {\hl -- see also Supplementary Figure S3b}.  Of course, if sufficiently high levels of behaviour could somehow be achieved without being triggered by high infection prevalence (e.g.,~via an advertising campaign or other public health intervention), then the same beneficial outcome may occur with a lower value of $R_0$ (i.e.,~if $B(0)$ is above approximately $0.3$ in Figure \ref{fig:cases3}d then the disease dies out). Furthermore, if $R_0$ is too large, the disease-free equilibrium with high $B$ eventually loses stability and the system moves into parameter regime 10, which always results in endemicity. {\hl Although a stronger behavioural effect raises the value of $R_0$ at which this happens, it remains finite even for $q_c=1$}. We also note that this mechanism requires both {\hl a high social behaviour uptake rate ($\tau > 4$) and sufficiently strong behavioural effect $q_c$}; if either condition fails, higher $R_0$ unambiguously worsens outcomes, as in classical models.

{\hl We have analysed the bifurcation structure of the two-dimensional reduction of the contagiousness-modulated SIS model in Eqs. \eqref{eq:2Dsystem1}--\eqref{eq:2Dsystem2}, in which $S_B(t)=S(t)B(t)$, i.e. susceptibility and behaviour are independent characteristics. Because this is an attracting invariant manifold for the full contagiousness-modulated SIS model in Eqs. \eqref{eq:3D_system1}--\eqref{eq:3D_system3} with $q_s=0$, the full model and the reduced model have the same equilibria and stability properties. However, in bistable cases, it is possible that the asymptotic state for the full model depends on $S_B(0)$, as well as on $S(0)$ and $B(0)$. To investigate this, Supplementary Figure S4 shows the large-time limit of numerical solutions of the full model for three different ways of setting the initial condition for $S_B(0)$: (i) independence of $S$ and $B$ (as in the reduced model); (ii) behavers maximally concentrated in the susceptible population; (iii) behavers maximally concentrated in the infectious population.  This demonstrates that the dependence of the long-term state on $S(0)$ and $B(0)$ is largely insensitive to the initial value for $S_B(0)$. Thus, the qualitative results presented here do not rely on initialising the model with $S_B(0)=S(0)B(0)$.  }

\subsection{Susceptibility-modulated SIS model}

In the susceptibility-modulated model (i.e., {\hl Eqs \eqref{eq:3D_system1}--\eqref{eq:FOI2}} with $q_c=0$), disease state transitions depend on behaviour which means that this model does not admit a two-dimensional reduction and so is less amenable to analysis. Instead, we compare numerical solutions of the model to corresponding solutions of the contagiousness-modulated model.

Numerical time series solutions of the susceptibility-modulated model (Eqs. \eqref{eq:3D_system1}--\eqref{eq:3D_system3} with $q_c=0$) are shown in the right column of plots in Figures \ref{fig:cases1}--\ref{fig:cases2} using the same set of parameter combinations and initial conditions as for the contagiousness-modulated model, but with $q_c=0$ and $q_s>0$. Since the susceptibility-modulated model is a three-dimensional system, a third variable $S_B(t)/S(t)$ is shown in these plots (yellow curves), as well as $I(t)$ and $B(t)$. This variable represents the proportion of the susceptible population that is in the behaving state at time $t$, in contrast to $B(t)$ which is the proportion of the overall population in the behaving state. 

These results shows that, for each of the eight selected parameter combinations, the susceptibility-modulated model has the same qualitative dynamics as the contagiousness-modulated model. The quantitative solutions are slightly different, and this is evident from the fact that $S_B(t)/S(t)$ is close to, but slightly larger than $B(t)$, i.e.,~the prevalence of {\hl behavers} is higher among the susceptible class than in the population as a whole. This is due to selective depletion of the susceptible pool: non-behavers are infected and move out of the susceptible compartment at a higher rate than behavers do. The bifurcations separating parameter regimes shown in Figure \ref{fig:bifurcation} for the contagiousness-modulated model will be in slightly different places for the susceptibility-modulated model. Nevertheless, it is clear that the qualitative bifurcation structure of the susceptibility-modulated model is the same as that of the contagiousness-modulated model, in the case where {\hl the timescale for behaviour change $\alpha$ is $0.25$ and the rate of disease-induced behaviour uptake $\chi$ is $0.2$}. 

For these parameter values, the timescale for recovery from disease (1 time unit) is comparable in magnitude to the timescale for behaviour change ($1/\alpha=4$ time units). When the behavioural timescale is either shorter or longer than the disease timescale, the equilibria and bifurcation points are unchanged and the model dynamics are qualitatively very similar but on a faster or slower timescale. Supplementary Figures \ref{fig:supp1} and \ref{fig:supp2} show model results where $\alpha$ is increased or decreased by a factor of 10 respectively. Similarly, if the behaviour affects both contagiousness and susceptibility (i.e.,~$q_c=a$ and $q_s=b$ are both positive), the dynamics are qualitatively similar (although not quantitatively identical) to the contagiousness-modulated model with $q_c=1-(1-a)(1-b)$ and $q_s=0$ (Supplementary Figure \ref{fig:supp3} presents an example).

\subsection{SIRS model example}

The disease-free equilibria and their stability properties are the same for the SIRS model as for the SIS model. Therefore, any parameter combination with {\hl high social behaviour uptake rate} ($\tau>4$) and $R_0<R_{0,TC+}$ has a stable DFE with high {\hl prevalence of behavers}.

We do not attempt to comprehensively map the bifurcation structure of the endemic equilibria in this more general SIRS model. However, we demonstrate that there are parameter values for which the SIRS model also exhibits one of the key phenomena shown by the SIS model, namely an epidemic followed by extinction of the disease due to persistent high {\hl prevalence of behavers}. This distinguishes the model from the classical SIRS model in which the endemic equilibrium is globally asymptotically stable for $R_0>1$.

\begin{figure}
    \centering
    \includegraphics[width=\linewidth]{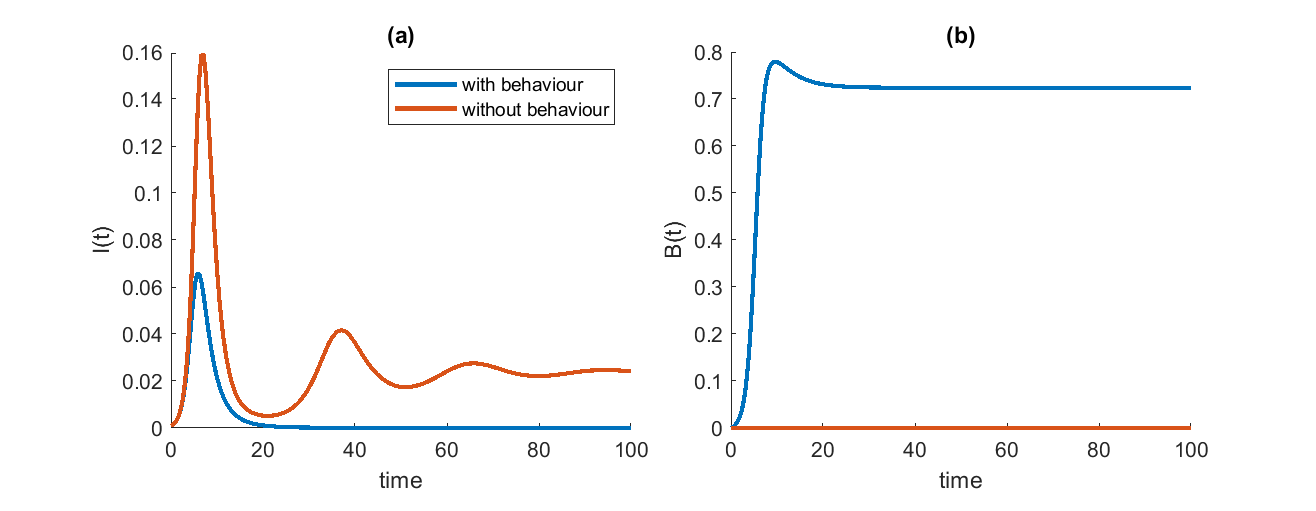}
    \caption{Solutions of the SIRS model showing: (a) prevalence of infection $I(t)=I_N(t)+I_B(t)$; (b) prevalence of behavers $B(t)=S_B(t)+I_B(t)+R_B(t)$. Solutions are shown with and without behavioural dynamics included in the model (blue and red curves respectively). Parameter values: $R_0=2$, $\nu=0.05$, $q_c=0.7$, $q_s=0$, $\tau=5$, $\chi=20$, $\alpha=0.25$. For the model without behavioural dynamics, $\chi$ and $\tau$ were set to zero, so that $B(t)=0$ for all $t$. Both solutions had the same initial condition: $S_N(0)=0.999$, $I_N(0)=0.001$ and all other compartments set equal to zero. }
    \label{fig:SIRS}
\end{figure}

Figure \ref{fig:SIRS} shows two solutions of the SIRS model, one with {\hl behavioural dynamics included in the model} and one in which $B(t)$ is set to $0$ for all $t$, for an example set of parameter values. Without behavioural dynamics, the model is simply the classical SIRS model which shows a series of epidemic waves of decreasing amplitude, tending towards the endemic equilibrium (Figure \ref{fig:SIRS}a, red). However, when behaviour is included in the model, the first epidemic wave is reduced in size, there are no subsequent waves, and the disease dies out (Figure \ref{fig:SIRS}a, blue). This occurs because the prevalence of behavers increases during the first wave and then remains at a high (and self-sustaining due to our model of complex contagion) level (Figure \ref{fig:SIRS}b, blue). This behaviour is the analogue of parameter regime 8 in the contagiousness-modulated SIS model: the system tends towards the BDFE$_+$, where the reproduction number $R^*_+$ (Eq. \eqref{eq:R_BDFE}) is less than $1$ so the disease cannot re-invade. The value of $\chi=20$ in Figure \ref{fig:SIRS} was chosen to ensure sufficient disease-induced behaviour uptake to produce the elimination outcome in the SIRS model; smaller values of $\chi$ produce qualitatively similar but quantitatively weaker behavioural responses.

\section{Discussion}

We have analysed a simplified model of infectious disease transmission in the presence of dynamic behavioural change, which reduces the susceptibility and/or contagiousness of those adopting the behaviour. Building on previous work, which assumed the rate of social behaviour uptake increases linearly with the number of people adopting the behaviour \cite{ryan2024behaviour,ryan2026behaviour}, we explored a `complex contagion' model for behavioural transmission, where this dependence is quadratic instead of linear. This models a situation where people are much more likely to adopt the behaviour if multiple of their contacts are also behaving. 

Our analysis reveals that the presence of complex contagion means that, in some situations, the occurrence of an epidemic drives an increase in behaviour that subsequently causes the disease to die out. For this mechanism to occur requires that behaviour uptake is sufficiently fast relative to behaviour abandonment and behaviour has a sufficiently strong impact on transmission. It also requires $R_0$ to be large enough that the disease initially becomes sufficiently prevalent to trigger widespread behaviour adoption, but not so large as to prevent elimination from occurring. Furthermore, behaviour-driven elimination can only occur if high prevalence of behaviour is self-sustaining {\hl as infection prevalence becomes low: if behaviour declines as infection prevalence drops, the epidemic may be able to spread again \cite{epstein2008coupled,perra2011towards}. This explains why models with simple behavioural contagion, despite the rich bifurcation structure and bistability they possess \cite{epstein2021triple,kaur2025impact}, do not exhibit this particular phenomenon.} If the behaviour uptake rate is linear in behaviour prevalence, there can only be one disease-free equilibrium that is stable to small perturbations in behaviour. Either the disease can invade from this equilibrium, in which case it will become endemic, or it cannot. 

When there is no infection-acquired immunity and behaviour only reduces contagiousness and not susceptibility, the model reduces to a two-dimensional dynamical system.  This reduction has recently been observed in a similar contagiousness-modulated model where behaviour is informed through imitation dynamics \cite{martin2026wearing}.
Reducing the system to two dimensions allows the bifurcation structure of the model to be comprehensively analysed, which provides a complete mechanistic understanding of the system.  The complex contagion in our model allows for a wider variety of bifurcations than those found by Martin \etal \cite{martin2026wearing}.  Additionally, we have shown that the same qualitative outcomes occur where behaviour affects both contagiousness and susceptibility, and in the more general case where there is some temporary post-recovery immunity.

There is evidence of persistent behaviour change driven by epidemics. For example, Salon \etal \cite{salon2021potential} found evidence that some behaviours induced by the Covid-19 pandemic, including working from home and reduced air travel, showed signs of persistence, though the extent of this varied considerably across behaviour types. The 2003 SARS epidemic led to a change in social norms for face mask usage across East Asia \cite{tsang2021boundaries}. During the epidemic, behaviour compliance became a symbol of social unity in controlling the outbreak. The memory of SARS has led to rapid uptake of face mask usage for respiratory outbreaks in areas affected by the initial infection \cite{chen2020impact}. Persistent change in other behavioural attributes following the SARS epidemic has also been observed \cite{lau2005impacts}. This is reminiscent of our results indicating a shift in attitudes and behavioural compliance after epidemic outbreaks, even when infection is eliminated. Sustained behaviour change has also been proposed as an important factor in the control of sexually transmitted infections \cite{vermund2021psychosocial,dewit2023sexually}. {\hl It is not realistic to assume that disease-induced behaviour change will always persist indefinitely. Nevertheless, provided it is sustained long enough, our results show that complex behavioural contagion can be sufficient to cause a qualitative shift from endemicity to elimination, even if lapse in behaviour subsequently leaves the population susceptible to re-introduction of the pathogen.  }

Our model does not account for all the psychological factors described by Ryan \etal \cite{ryan2024behaviour}.  In particular, it does not account for potential social contagion of `not behaving' or avenues for spontaneous behaviour uptake. In the model of Ryan \etal \cite{ryan2024behaviour}, social contagion of non-behaviour can cause infection to persist due to non-behavers sustaining transmission.  

An important practical implication of our analysis concerns the parameter $\tau$, which governs the rate of social behaviour uptake relative to abandonment. Our results show that $\tau=4$ is a critical threshold: below it, the only disease-free equilibrium has zero behaviour, whereas above it, self-sustaining high-behaviour equilibria emerge. In the Health Belief Model framework of Ryan \etal \cite{ryan2024behaviour}, $\tau$ maps broadly onto the balance between social cues to action, captured by the parameter governing the infectiousness of visible behaviour (their $\omega_1$), and the combined forces driving behaviour abandonment. {\hl The behavioural effect $q_c$ sets how much a given level of behaviour reduces contagiousness, but not the level of behaviour the population reaches, which in the disease-free system depends on $\tau$ alone. The largest $R_0$ that behaviour can hold at bay is $1/B^*_-$, the reciprocal of the behavioural tipping threshold, and lowering that threshold requires strengthening the social process rather than making the behaviour more protective.} This suggests that public health interventions aimed at strengthening social norms around protective behaviour, for example by increasing the visibility of compliant behaviour or reducing the social acceptability of abandoning it, could shift a population across this threshold and qualitatively improve the long-term disease outcome.

There are several ways our model could be generalised. We have explored a situation where behaviour reduces disease transmission by reducing susceptibility and/or contagiousness. An analogous situation that could be relevant in some applications is adoption of a risky behaviour that increases susceptibility and/or contagiousness. This could be investigated within our framework, but would require a more general model for the behaviour uptake and abandonment rates. Our model assumed that behaviour uptake increases with disease prevalence, representing more precautionary behaviour in response to perceived risk. When modelling a risky behaviour as opposed to a precautionary one, it would be more realistic to assume that, as disease prevalence increases, behaviour uptake decreases and/or behaviour abandonment increases. 

Other generalisations of the behaviour uptake and abandonment model are also possible and may be relevant in applications. For example, the uptake and abandonment rates could be general polynomials with a mixture of linear and nonlinear terms in both the prevalence of {\hl behavers} $B$ and the prevalence of disease $I$ \cite{kastalskiy2021social}. 
Furthermore, our model and that of \cite{ryan2024behaviour} assume that behavioural change is independent of disease state. This may not always be true as, for example, infectious individuals may be more (or less) likely than susceptible individuals to adopt the behaviour depending on the context. Recovered individuals may be less likely to adopt the behaviour because there is no benefit of doing so, either to the individual or to the population. {\hl Game theoretic approaches, which have been previously applied to modelling vaccination decisions \cite{bauch2004vaccination,oraby2014influence,morsky2025vaccination} and other behavioural choices \cite{gersovitz2004economical,reluga2010game,fenichel2011adaptive,plank2026joint}, may be useful to address how people weigh different behavioural drivers (e.g. risk perception and social contagion) and the dependence of behaviour on disease state.}

\subsection*{Data availability statement}
No new data is presented in this article. The code that was used to produce the results in the article is publicly available at \cite{github-repo}. 

\subsection*{Competing interest statement}
The authors declare no competing interests.

\subsection*{Funding statement}
MJP was supported by the Royal Society Te Ap\=arangi Marsden Fund (24-UOC-020) and Te Niwha Infectious Diseases Research Platform, co-hosted by PHF Science and the University of Otago and provisioned by the Ministry of Business, Innovation and Employment, New Zealand (TN/P/24/UoC/MP). JMM was supported by an Australian Research Council Laureate Fellowship (FL240100126). This work was supported by EPSRC grant no EP/K032208/1.

\subsection*{Acknowledgements}
The authors would like to thank the Isaac Newton Institute for Mathematical Sciences, Cambridge, for support and hospitality during the programme Modelling and inference for pandemic preparedness where work on this paper was undertaken. This work was supported by EPSRC grant no EP/K032208/1. The authors are grateful to Hans Heesterbeek, Valerie Isham, Emily Nixon and Mick Roberts for helpful discussions and feedback on the model and to two anonymous peer reviewers for comments on an earlier version of this manuscript.

\bibliography{references}

@article{ryan2024behaviour,
  title={A behaviour and disease transmission model: incorporating the Health Belief Model for human behaviour into a simple transmission model},
  author={Ryan, Matthew and Brindal, Emily and Roberts, Mick and Hickson, Roslyn I},
  journal={Journal of the Royal Society Interface},
  volume={21},
  number={215},
  pages={20240038},
  year={2024},
  publisher={The Royal Society}
}

@article{ryan2026behaviour,
  title={A behaviour and disease model of testing and isolation},
  author={Ryan, Matthew and Hickson, Roslyn I and Hill, Edward M and House, Thomas and Isham, Valerie and Zhang, Dongni and Roberts, Mick G},
  journal={Mathematics in Medical and Life Sciences},
  year={2026},
  volume = {3},
  pages = {10.1080/29937574.2026.2627906}
}

@misc{github-repo,
   title = {Github repository: Behaviour and disease SIS model},
   author = {Plank, M J and McCaw, J},
   year = {2026},
   url = {https://github.com/michaelplanknz/behaviour_and_disease_SIS_model}
}

@article{marion2022modelling,
  title={Modelling: understanding pandemics and how to control them},
  author={Marion, Glenn and Hadley, Liza and Isham, Valerie and Mollison, Denis and Panovska-Griffiths, Jasmina and Pellis, Lorenzo and Tomba, Gianpaolo Scalia and Scarabel, Francesca and Swallow, Ben and Trapman, Pieter and others},
  journal={Epidemics},
  volume={39},
  pages={100588},
  year={2022},
  publisher={Elsevier}
}

@article{bedson2021review,
  title={A review and agenda for integrated disease models including social and behavioural factors},
  author={Bedson, Jamie and Skrip, Laura A and Pedi, Danielle and Abramowitz, Sharon and Carter, Simone and Jalloh, Mohamed F and Funk, Sebastian and Gobat, Nina and Giles-Vernick, Tamara and Chowell, Gerardo and others},
  journal={Nature Human Behaviour},
  volume={5},
  number={7},
  pages={834--846},
  year={2021},
  publisher={Nature Publishing Group UK London}
}

@article{funk2010modelling,
  title={Modelling the influence of human behaviour on the spread of infectious diseases: a review},
  author={Funk, Sebastian and Salath{\'e}, Marcel and Jansen, Vincent AA},
  journal={Journal of the Royal Society Interface},
  volume={7},
  number={50},
  pages={1247--1256},
  year={2010},
  publisher={The Royal Society}
}

@article{sprague2017evidence,
  title={Evidence for complex contagion models of social contagion from observational data},
  author={Sprague, Daniel A and House, Thomas},
  journal={PLoS One},
  volume={12},
  number={7},
  pages={e0180802},
  year={2017},
  publisher={Public Library of Science San Francisco, CA USA}
}

@article{centola2007complex,
  title={Complex contagions and the weakness of long ties},
  author={Centola, Damon and Macy, Michael},
  journal={American Journal of Sociology},
  volume={113},
  number={3},
  pages={702--734},
  year={2007},
  publisher={The University of Chicago Press}
}

@article{centola2010spread,
  title={The spread of behavior in an online social network experiment},
  author={Centola, Damon},
  journal={Science},
  volume={329},
  number={5996},
  pages={1194--1197},
  year={2010},
  publisher={American Association for the Advancement of Science}
}

@article{kastalskiy2021social,
  title={Social stress drives the multi-wave dynamics of COVID-19 outbreaks},
  author={Kastalskiy, Innokentiy A and Pankratova, Evgeniya V and Mirkes, Evgeny M and Kazantsev, Victor B and Gorban, Alexander N},
  journal={Scientific Reports},
  volume={11},
  number={1},
  pages={22497},
  year={2021},
  publisher={Nature Publishing Group UK London}
}

@article{funk2009spread,
  title={The spread of awareness and its impact on epidemic outbreaks},
  author={Funk, Sebastian and Gilad, Erez and Watkins, Chris and Jansen, Vincent A A},
  journal={Proceedings of the National Academy of Sciences},
  volume={106},
  number={16},
  pages={6872--6877},
  year={2009},
  publisher={National Academy of Sciences}
}

@article{chang2025impact,
  title={Impact of opinion dynamics on recurrent pandemic waves: balancing risk aversion and peer pressure},
  author={Chang, Sheryl L and Nguyen, Quang Dang and Suster, Carl Joseph Edmund and Jamerlan, Ma Christina and Rockett, Rebecca J and Sintchenko, Vitali and Sorrell, Tania C and Martiniuk, Alexandra and Prokopenko, Mikhail},
  journal={Interface Focus},
  volume={15},
  number={4},
  pages={20240038},
  year={2025},
  publisher={The Royal Society}
}

@article{auld2025economics,
  title={The Economics of Infectious Diseases},
  author={Auld, M Christopher and Fenichel, Eli P and Toxvaerd, Flavio},
  journal={Journal of Economic Literature},
  year={2025},
  volume = {63},
  pages = {1281-1330}
}

@article{hill2024integrating,
  title={Integrating human behaviour and epidemiological modelling: unlocking the remaining challenges},
  author={Hill, Edward M and Ryan, Matthew and Haw, David and Lynch, Mark P and McCabe, Ruth and Milne, Alice E and Turner, Matthew S and Vedhara, Kavita and Zeng, Fanqi and Barons, Martine J and others},
  journal={Mathematics in Medical and Life Sciences},
  volume={1},
  number={1},
  pages={2429479},
  year={2024},
  publisher={Taylor \& Francis}
}

@article{rosenstock1966people,
  title={Why people use health services},
  author={Rosenstock, Irwin M},
  journal={{Milbank Memorial Fund Quarterly}},
  volume={44},
  number={3},
  pages={94--127},
  year={1966},
  publisher={JSTOR}
}

@article{becker1974health,
  title={The health belief model and sick role behavior},
  author={Becker, Marshall H},
  journal={{Health Education Monographs}},
  volume={2},
  number={4},
  pages={409--419},
  year={1974},
  publisher={SAGE Publications Sage CA: Los Angeles, CA}
}

@article{martin2026wearing,
  title={Wearing face masks to protect oneself and/or others: counter-intuitive results from a simple epidemic model accounting for selfish and altruistic human behavior},
  author={Martin, Hugo and Castella, Fran{\c{c}}ois and Hamelin, Fr{\'e}d{\'e}ric},
  journal={Journal of Theoretical Biology},
  pages={112395},
  year={2026},
  publisher={Elsevier}
}

@article{chen2020impact,
  title={The impact of social ties and SARS memory on the public awareness of 2019 novel coronavirus (SARS-CoV-2) outbreak},
  author={Chen, Haohui and Paris, Cecile and Reeson, Andrew},
  journal={Scientific Reports},
  volume={10},
  number={1},
  pages={18241},
  year={2020},
  publisher={Nature Publishing Group UK London}
}

@article{tsang2021boundaries,
  title={Boundaries of solidarity: a meta-ethnography of mask use during past epidemics to inform SARS-CoV-2 suppression},
  author={Tsang, Po Man and Prost, Audrey},
  journal={BMJ Global Health},
  volume={6},
  number={1},
  year={2021},
  publisher={BMJ Publishing Group Ltd}
}

@article{salon2021potential,
  title={The potential stickiness of pandemic-induced behavior changes in the United States},
  author={Salon, Deborah and Conway, Matthew Wigginton and Capasso da Silva, Denise and Chauhan, Rishabh Singh and Derrible, Sybil and Mohammadian, Abolfazl and Khoeini, Sara and Parker, Nathan and Mirtich, Laura and Shamshiripour, Ali and others},
  journal={Proceedings of the National Academy of Sciences},
  volume={118},
  number={27},
  pages={e2106499118},
  year={2021},
  publisher={National Academy of Sciences}
}

@incollection{vermund2021psychosocial,
  title={Psychosocial and behavioral interventions},
  author={Vermund, Sten H and Geller, Amy B and Crowley, Jeffrey S and {Committee on Prevention and Control of Sexually Transmitted Infections in the United States} and {Board on Population Health and Public Health Practice; Health and Medicine Division} and {National Academies of Sciences, Engineering, and Medicine}},
  booktitle={Sexually Transmitted Infections: Adopting a Sexual Health Paradigm},
  year={2021},
  pages={399-462},
  publisher={National Academies Press}
}

@article{dewit2023sexually,
  title={Sexually transmitted infection prevention behaviours: health impact, prevalence, correlates, and interventions},
  author={de Wit, John B F and Adam, Philippe C G and den Daas, Chantal and Jonas, Kai},
  journal={Psychology and Health},
  volume={38},
  number={6},
  pages={675--700},
  year={2023},
  publisher={Taylor \& Francis}
}

@article{raude2019understanding,
  title={Understanding health behaviour changes in response to outbreaks: Findings from a longitudinal study of a large epidemic of mosquito-borne disease},
  author={Raude, Jocelyn and McColl, Kathleen and Flamand, Claude and Apostolidis, Themis},
  journal={Social Science and Medicine},
  volume={230},
  pages={184--193},
  year={2019},
  publisher={Elsevier}
}

@article{gimma2022changes,
  title={Changes in social contacts in England during the COVID-19 pandemic between March 2020 and March 2021 as measured by the CoMix survey: A repeated cross-sectional study},
  author={Gimma, Amy and Munday, James D and Wong, Kerry L M and Coletti, Pietro and van Zandvoort, Kevin and Prem, Kiesha and {CMMID COVID-19 working group} and Klepac, Petra and Rubin, G James and Funk, Sebastian and others},
  journal={PLoS Medicine},
  volume={19},
  number={3},
  pages={e1003907},
  year={2022},
  publisher={Public Library of Science San Francisco, CA USA}
}

@article{eales2025temporal,
  title={Temporal trends in test-seeking behaviour during the COVID-19 pandemic},
  author={Eales, Oliver and Teo, Mingmei and Price, David J and Hao, Tianxiao and Ryan, Gerard E and Senior, Katharine L and Carlson, Sandra and Dalton, Craig and Dawson, Peter and Golding, Nick and others},
  journal={Mathematics in Medical and Life Sciences},
  volume={2},
  number={1},
  pages={2521858},
  year={2025},
  publisher={Taylor \& Francis}
}

@article{lau2005impacts,
  title={Impacts of SARS on health-seeking behaviors in general population in Hong Kong},
  author={Lau, Joseph T F and Yang, Xilin and Tsui, H Y and Kim, Jean H},
  journal={Preventive Medicine},
  volume={41},
  number={2},
  pages={454--462},
  year={2005},
  publisher={Elsevier}
}

@article{tang2004factors,
  title={Factors influencing the wearing of facemasks to prevent the severe acute respiratory syndrome among adult Chinese in Hong Kong},
  author={Tang, Catherine So-kum and Wong, Chi-yan},
  journal={Preventive Medicine},
  volume={39},
  number={6},
  pages={1187--1193},
  year={2004},
  publisher={Elsevier}
}

@article{joffe2003risk,
  title={Risk: From perception to social representation},
  author={Joffe, H{\'e}l{\`e}ne},
  journal={British Journal of Social Psychology},
  volume={42},
  number={1},
  pages={55--73},
  year={2003},
  publisher={Wiley Online Library}
}

@article{morsky2025vaccination,
  title={Vaccination and collective action under social norms},
  author={Morsky, Bryce},
  journal={Bulletin of Mathematical Biology},
  volume={87},
  number={5},
  pages={55},
  year={2025},
  publisher={Springer}
}

@article{oraby2014influence,
  title={The influence of social norms on the dynamics of vaccinating behaviour for paediatric infectious diseases},
  author={Oraby, Tamer and Thampi, Vivek and Bauch, Chris T},
  journal={Proceedings of the Royal Society B: Biological Sciences},
  volume={281},
  number={1780},
  pages={20133172},
  year={2014}
}

@article{contreras2024infection,
  title={Infection patterns in simple and complex contagion processes on networks},
  author={Contreras, Diego Andr{\'e}s and Cencetti, Giulia and Barrat, Alain},
  journal={PLoS Computational Biology},
  volume={20},
  number={6},
  pages={e1012206},
  year={2024},
  publisher={Public Library of Science, San Francisco, CA USA}
}

@article{campbell2013complex,
  title={Complex social contagion makes networks more vulnerable to disease outbreaks},
  author={Campbell, Ellsworth and Salath{\'e}, Marcel},
  journal={Scientific Reports},
  volume={3},
  number={1},
  pages={1905},
  year={2013},
  publisher={Nature Publishing Group UK, London}
}

@article{bauch2012evolutionary,
  title={Evolutionary game theory and social learning can determine how vaccine scares unfold},
  author={Bauch, Chris T and Bhattacharyya, Samit},
  journal={PLoS Computational Biology},
  volume={8},
  number={4},
  pages={e1002452},
  year={2012},
  publisher={Public Library of Science, San Francisco, USA}
}

@article{bauch2004vaccination,
  title={Vaccination and the theory of games},
  author={Bauch, Chris T and Earn, David JD},
  journal={Proceedings of the National Academy of Sciences},
  volume={101},
  number={36},
  pages={13391--13394},
  year={2004},
  publisher={National Academy of Sciences}
}

@article{epstein2021triple,
  title={Triple contagion: a two-fears epidemic model},
  author={Epstein, Joshua M and Hatna, Erez and Crodelle, Jennifer},
  journal={Journal of the Royal Society Interface},
  volume={18},
  number={181},
  pages={20210186},
  year={2021}
}

@article{kaur2025impact,
  title={The impact of fear and behavior response to established and novel diseases},
  author={Kaur, Avneet and Tyson, Rebecca C and Moyles, Iain R},
  journal={SIAM Journal on Applied Mathematics},
  volume={85},
  number={2},
  pages={687--710},
  year={2025},
  publisher={SIAM}
}

@article{epstein2008coupled,
  title={Coupled contagion dynamics of fear and disease: mathematical and computational explorations},
  author={Epstein, Joshua M and Parker, Jon and Cummings, Derek and Hammond, Ross A},
  journal={PLoS One},
  volume={3},
  number={12},
  pages={e3955},
  year={2008},
  publisher={Public Library of Science, San Francisco, USA}
}

@article{perra2011towards,
  title={Towards a characterization of behavior-disease models},
  author={Perra, Nicola and Balcan, Duygu and Gon{\c{c}}alves, Bruno and Vespignani, Alessandro},
  journal={PLoS One},
  volume={6},
  number={8},
  pages={e23084},
  year={2011},
  publisher={Public Library of Science, San Francisco, USA}
}

@article{plank2026joint,
  title={Joint economic and epidemiological modelling of alternative pandemic response strategies},
  author={Plank, Michael John and Sushames, Megan and Fisher-Taylor, Timothy and Afshari, Amin and Thompson, Robin N and Hurford, Amy and Hendy, Shaun},
  journal={Journal of the Royal Society Interface},
  volume={23},
  number={241},
  year={2026},
  publisher={The Royal Society}
}

@article{reluga2010game,
  title={Game theory of social distancing in response to an epidemic},
  author={Reluga, Timothy C},
  journal={PLoS Computational Biology},
  volume={6},
  number={5},
  pages={e1000793},
  year={2010},
  publisher={Public Library of Science, San Francisco, USA}
}

@article{fenichel2011adaptive,
  title={Adaptive human behavior in epidemiological models},
  author={Fenichel, Eli P and Castillo-Chavez, Carlos and Ceddia, M Graziano and Chowell, Gerardo and Parra, Paula A Gonzalez and Hickling, Graham J and Holloway, Garth and Horan, Richard and Morin, Benjamin and Perrings, Charles and others},
  journal={Proceedings of the National Academy of Sciences},
  volume={108},
  number={15},
  pages={6306--6311},
  year={2011},
  publisher={National Academy of Sciences}
}

@article{gersovitz2004economical,
  title={The economical control of infectious diseases},
  author={Gersovitz, Mark and Hammer, Jeffrey S},
  journal={The Economic Journal},
  volume={114},
  number={492},
  pages={1--27},
  year={2004},
  publisher={Oxford University Press, Oxford, UK}
}

@article{tyson2022transient,
  title={Transient prophylaxis and multiple epidemic waves},
  author={Tyson, Rebecca C and Marshall, Noah D and Baumgaertner, Bert O},
  journal={AIMS Mathematics},
  volume={7},
  number={4},
  pages={5616--5633},
  year={2022}
}

\clearpage

\resetlinenumber

\renewcommand\theequation{S\arabic{equation}}
\renewcommand\thefigure{S\arabic{figure}}
\renewcommand\thetable{S\arabic{table}}
\renewcommand\thesection{S\arabic{section}}

\setcounter{equation}{0}
\setcounter{figure}{0}
\setcounter{table}{0}
\setcounter{section}{0}
\pagenumbering{arabic}

{\huge How complex behavioural contagion can prevent infectious diseases from becoming endemic: Supplementary Material}

\section{SIRS model equations}

The SIRS model is described by the following system of ordinary differential equations 
\begin{align} 
\frac{dS_N}{dt} &= -\lambda S_N + \nu R_N - \omega(I,B) S_N + \alpha S_B \label{eq:SIRS1} \\
\frac{dI_N}{dt} &= \lambda S_N -  I_N - \omega(I,B) I_N + \alpha I_B \\
\frac{dR_N}{dt} &= I_N - \nu R_N - \omega(I,B) R_N + \alpha R_B \\
\frac{dS_B}{dt} &= - (1-q_s) \lambda S_B +  \nu R_B + \omega(I,B) S_N - \alpha S_B \\
\frac{dI_B}{dt} &=  (1-q_s) \lambda S_B -  I_B + \omega(I,B) I_N - \alpha I_B \\
\frac{dR_B}{dt} &=  I_B - \nu R_B + \omega(I,B) R_N - \alpha R_B \label{eq:SIRS6}
\end{align}
where the force of infection $\lambda$ is defined as before by Eq. (5) and the behaviour uptake rate $\omega$ is given by 
\begin{equation}
\omega(I,B) = \alpha\left( \tau B^2 + \chi I \right)
\end{equation}
Note that the SIS model in Eqs. (1)--(4) is the limiting case of this SIRS model when the immune waning parameter $\nu$ tends to $\infty$.

\section{Contagiousness-modulated SIS model} 

\subsection{Model reduction}
\label{sec:model_reduction}

To reduce the contagiousness-modulated SIS model to a two-dimensional system, we first express Eqs. \eqref{eq:3D_system1}--\eqref{eq:3D_system3} with $q_s=0$ in terms of $S=S_N+S_B$, $B=S_B+I_B$ and $S_B$:
\begin{align}
\frac{dS}{dt} &= -R_0 S \left(1-S - q_c (B-S_B)\right) +  1-S \label{eq:dSdt} \\
\frac{dB}{dt} &= \omega (1-B) - \alpha B, \\
\frac{dS_B}{dt} &= -R_0 S_B \left(1-S - q_c (B-S_B)\right) + B-S_B + \omega (S-S_B) - \alpha S_B \label{eq:dSBdt}
\end{align}

Then it follows that
\begin{align} \label{eq:SBdot}
\dot{S}B + S\dot{B} &=  -R_0 SB \left(1-S - q_c (B-S_B)\right) +  B(1-S) + \omega S(1-B) - \alpha SB
\end{align}
If $S_B=SB$, then Eq. \eqref{eq:SBdot} becomes
\begin{align} 
\dot{S}B + S\dot{B} &=  -R_0 SB (1-S)(1-q_cB) +  B(1-S) + \omega S(1-B) - \alpha SB
\end{align}
and Eq. \eqref{eq:dSBdt} becomes
\begin{align}
\dot{S_B} &= -R_0 SB(1-S) (1-q_c B) +  B(1-S) + \omega S(1-B) - \alpha SB  
\end{align}
which implies that $\dot{S_B}=d/dt(SB)$. Hence in the special case where the initial condition is such that $S_B(0)=S(0)B(0)$, then $S_B(t)$ must equal $S(t)B(t)$ for all $t>0$ as required. 

Furthermore, in the general case where $S_B(0)\neq S(0)B(0)$, subtracting Eq. \eqref{eq:SBdot} from Eq. \eqref{eq:dSBdt} gives 
\begin{align}
d/dt ( S_B - SB ) &=  -R_0 S_B \left(1-S - q_c (B-S_B)\right)+  (B-S_B) + \omega (S-S_B) - \alpha S_B \nonumber \\
& +R_0 SB\left(1-S - q_c (B-S_B)\right) -  B(1-S) - \omega S(1-B) + \alpha SB 
\end{align}

Writing $x(t) = S_B(t) - S(t)B(t)$, this reduces to
\begin{equation} \label{eq:xdot}
\dot{x} = -R_0 x\left(1-S - q_c (B-S_B)\right) -  x - \omega x - \alpha x 
\end{equation}
Each term on the right-hand side is a negative multiple of $x$ and therefore $x(t)$ decays monotonically to $0$ as $t$ increases. Hence, even if the initial condition does not allow reduction to a two-dimensional system, the dynamics will asymptotically relax onto the invariant two-dimensional manifold defined by $S_B = SB$. 

On substituting $S_B = SB$ into Eqs. \eqref{eq:dSdt}--\eqref{eq:dSBdt} and using the definition of $\omega$ in Eq. \eqref{eq:omega}, we obtain the two-dimensional system seen in Eqs. \eqref{eq:2Dsystem1}--\eqref{eq:2Dsystem2}.

Note that the first two terms in Eq. \eqref{eq:xdot} correspond to infection and recovery transitions respectively, while the last two correspond to behaviour uptake and behaviour abandonment respectively. This is because both disease state transitions and behaviour state transitions act to move the system towards the $x=0$ manifold. Hence the timescale for relaxation onto the manifold is determined by the faster of the disease dynamics timescale and the behaviour dynamics timescale.

\subsection{Equilibrium conditions} \label{sec:equilibrium_conditions}

The nullclines of the two-dimensional system defined by Eqs. \eqref{eq:2Dsystem1}--\eqref{eq:2Dsystem2} are
\begin{align}
S^* &= \frac{1}{R_0(1 - q_c B^*)} \label{eq:Snull}  \\
B^* &= (1-B^*)\left[\tau B^{*2} + \chi(1-S^*)\right] \label{eq:Bnull}
\end{align}

Therefore, any endemic equilibria $(S^*,B^*)$ must satisfy
\begin{equation} \label{eq:EE}
B^*(1 -q_c B^*) = (1-B^*)\left[ \tau B^{*2}(1 -q_c B^*) + \chi \left(1 -q_c B^*-\frac{1}{R_0}\right)\right] 
\end{equation}
which is a polynomial in $B^*$. Saddle node bifurcations occur when there is a repeated root of this polynomial, which requires the derivative of the polynomial with respect to $B^*$ to be zero. Hence saddle node bifurcations must also satisfy
\begin{align} \label{eq:SNB}
 (1-2q_c B^*) &= \tau \left[ 2 B^{*} - 3(1+q_c)B^{*2} + 4q_c B^{*3} \right]  +  \chi \left[ -q_c + 2q_cB^* + \frac{1}{R_0} -1    \right] 
\end{align}
For given values of $\tau$, $\chi$ and $q_c$, we solve Eqs. \eqref{eq:EE}--\eqref{eq:SNB} numerically to find the values of $R_0$ and $B^*$ at SNB2 and (if $\tau<4$) SNB1.

\subsection{Bifurcation structure} \label{sec:bifurcation_analysis}

Here we describe the bifurcations of the two-dimensional contagiousness-modulated SIS model and the resulting changes in dynamics that occur as $R_0$ increases for a fixed value of $q_c$, i.e.,~as parameters move from left to right along a horizontal line in either Figure \ref{fig:bifurcation}a or \ref{fig:bifurcation}b.

\subsubsection*{Low social behaviour uptake rate}
When $\tau<4$ (Figure \ref{fig:bifurcation}a), the only disease-free equilibrium has $B=0$ (NDFE). In this case, as in classical models, there is a transcritical bifurcation at $R_0=1$ of the NDFE and an endemic equilibrium (labelled TC0 in Figure \ref{fig:bifurcation}). When $R_0$ is less than 1, the disease cannot invade and the NDFE is globally stable (regime 1). As $R_0$ increases above $1$, an endemic equilibrium moves into the biological region (i.e.,~$(S,B)\in[0,1]^2$) and becomes stable (regime 2, Figure \ref{fig:cases1}a,b).

As $R_0$ increases, the $S$ nullcline moves to the left (i.e.,~the herd immunity threshold increases) and eventually intersects the upper part of the sigmoidally-shaped $B$ nullcline. This causes a saddle node bifurcation (SNB1), which generates two additional endemic equilibria (one of which is stable, while the other one is a saddle), which have higher $B$ (and lower $S$) than the existing endemic equilibrium  (regime 3, Figure \ref{fig:cases1}d,e). In this regime, the system has two stable endemic equilibria, whose basins of attraction are separated by the stable manifold of the saddle endemic equilibrium (light blue curve in Figure \ref{fig:cases1}d). The unstable manifold consists of two heteroclinic orbits connecting the saddle equilibrium to the two stable equilibria (dark red curve, Figure \ref{fig:cases1}d). 

As $R_0$ increases further, the saddle equilibrium moves towards the endemic equilibrium with low $B$ and eventually annihilates it in a second saddle node bifurcation (SNB2). Thus the only remaining stable equilibrium is the endemic equilibrium with high $B$ (regime 4, Figure \ref{fig:cases1}g,h).

An example of this sequence of regime changes as $R_0$ increases is shown in Supplementary Figure \ref{fig:prevalence_vs_R0}a for $\tau=3.8$ and $q_c=0.3$.

\subsubsection*{High social behaviour uptake rate}
 When $\tau>4$ (Figure \ref{fig:bifurcation}b), there are three disease-free equilibria: one with $B^*=0$ (NDFE) and two with $B^*>0$ (BDFE$_\pm$). As $R_0$ increases through one, a stable endemic equilbrium again enters the biological region via a transcritical bifurcation with the NDFE. However, when $\tau>4$, BDFE$_+$ is also stable, which means that the system is bistable. If $B(0)$ is sufficiently high, the system tends towards BDFE$_+$ and the disease ultimately dies out (regimes 6 and 7, Figure \ref{fig:cases3}a,b and d,e). Note the only difference between regimes 6 and 7 is that in regime 6, BDFE$_-$ is a saddle and there is only one endemic equilibrium; whereas in regime 7, BDFE$_-$ is an unstable node, and there is a second endemic equilibrium, which is a saddle. 
 The asymptotic behaviour is qualitatively the same in both regimes: endemicity if the initial condition is below the stable manifold of the saddle equilibrium (light blue curve, Figure \ref{fig:cases3}a and d), or extinction at BDFE$_+$. However, regime 7 admits solutions where there is a transient outbreak followed by extinction, while in regime 6 all initial conditions in the basin of attraction of BDFE$_+$ have monotonically decaying prevalence.

The effect of further increasing $R_0$ depends on the strength of the behavioural effect $q_c$. If the behavioural effect is strong, then as $R_0$ increases, the saddle node bifurcation that destroys the endemic equilibrium with low $B^*$ (SNB2) occurs before the transcritical bifurcation in which BDFE$_+$ loses stability (TC$+$). Hence, there are values of $R_0$ for which no stable endemic equilibrium exists, and the only stable equilibrium is BDFE$_+$. In these cases, all initial conditions lead to a transient epidemic, followed by the disease dying out and behaviour persisting (regime 8, Figure \ref{fig:cases3}g,h).  
Further increases in $R_0$ eventually lead to transcritical bifurcation of the BDFE$_+$ (TC$+$), meaning that all initial conditions tend to the endemic equilibrium with high $B^*$ (regime 10, \ref{fig:cases2}d,e). 

An example of this sequence of regime changes as $R_0$ increases is shown in Supplementary Figure \ref{fig:prevalence_vs_R0}b for $\tau=5$ and $q_c=0.7$.

If the behavioural effect is weak, then as $R_0$ increases, TC$+$ occurs before SNB2. TC$+$ leads to a second stable endemic equilibrium with high $B^*$ entering the biological region as BDFE$_+$ loses stability. All initial conditions now tend to one of the two stable endemic equilibria (regime 9, Figure \ref{fig:cases2}a,b).
Further increases in $R_0$ eventually lead to destruction of the stable endemic equilibrium with low $B^*$ via SNB2, meaning that all initial conditions tend to the endemic equilibrium with high $B^*$ (regime 10, Figure \ref{fig:cases2}d,e). 

An example of this sequence of regime changes as $R_0$ increases is shown in Supplementary Figure \ref{fig:prevalence_vs_R0}c for $\tau=5$ and $q_c=0.3$.

\section{Supplementary figures}

\begin{figure}
    \centering
    \includegraphics[trim={0.5cm 0cm 0.5cm 0cm},clip,width=\linewidth]{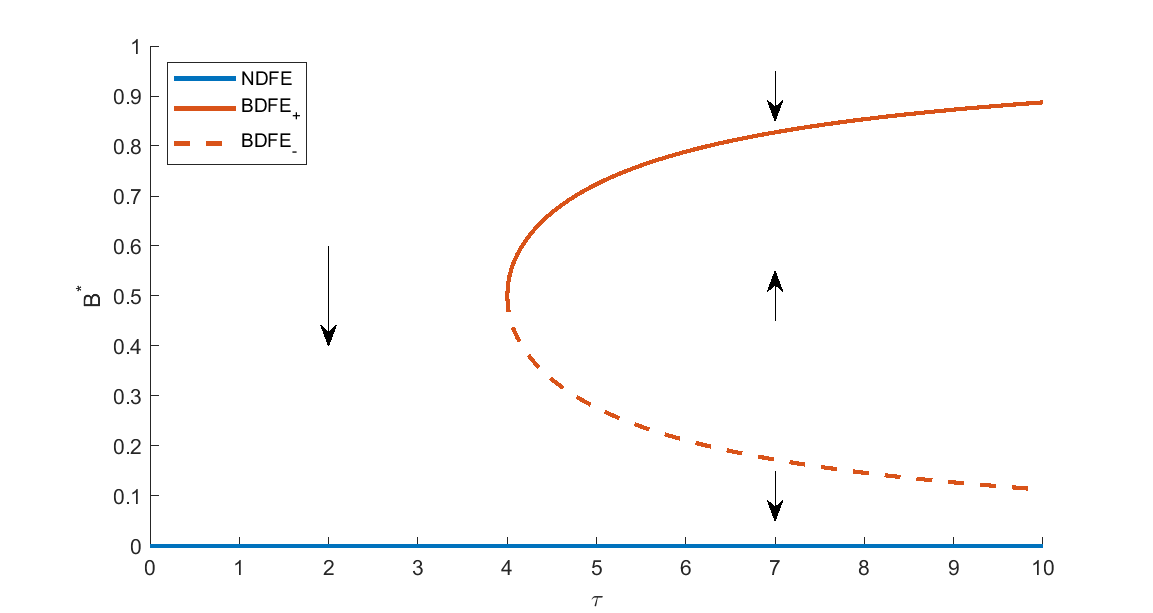}
    \caption{Bifurcation diagram for the invariant disease-free subsystem of the model showing the equilibrium prevalence of behavers $B^*$ against the parameter $\tau$, representing the rate of social behaviour uptake relative to behaviour abandonment. The disease-free equilibrium with $B=0$ (NDFE, blue) is always locally stable. Two disease-free equilibria with behaviour (red) are created via a saddle node bifurcation at $\tau=4$. The equilibrium with high $B$ (BDFE$_+$, red solid) is stable and the equilibrium with low $B$ (BDFE$_-$, red dashed) is unstable. Black arrows show the direction of change of $B(t)$. }
    \label{fig:disease_free}
\end{figure}

Supplementary Figure \ref{fig:disease_free} shows the bifurcation diagram for the invariant disease-free subsystem of the model with respect to the parameter $\tau$, representing the rate of social behaviour uptake relative to behaviour abandonment.

\begin{figure}
    \centering
    \includegraphics[trim={0.5cm 0cm 0.5cm 0cm},clip,width=\linewidth]{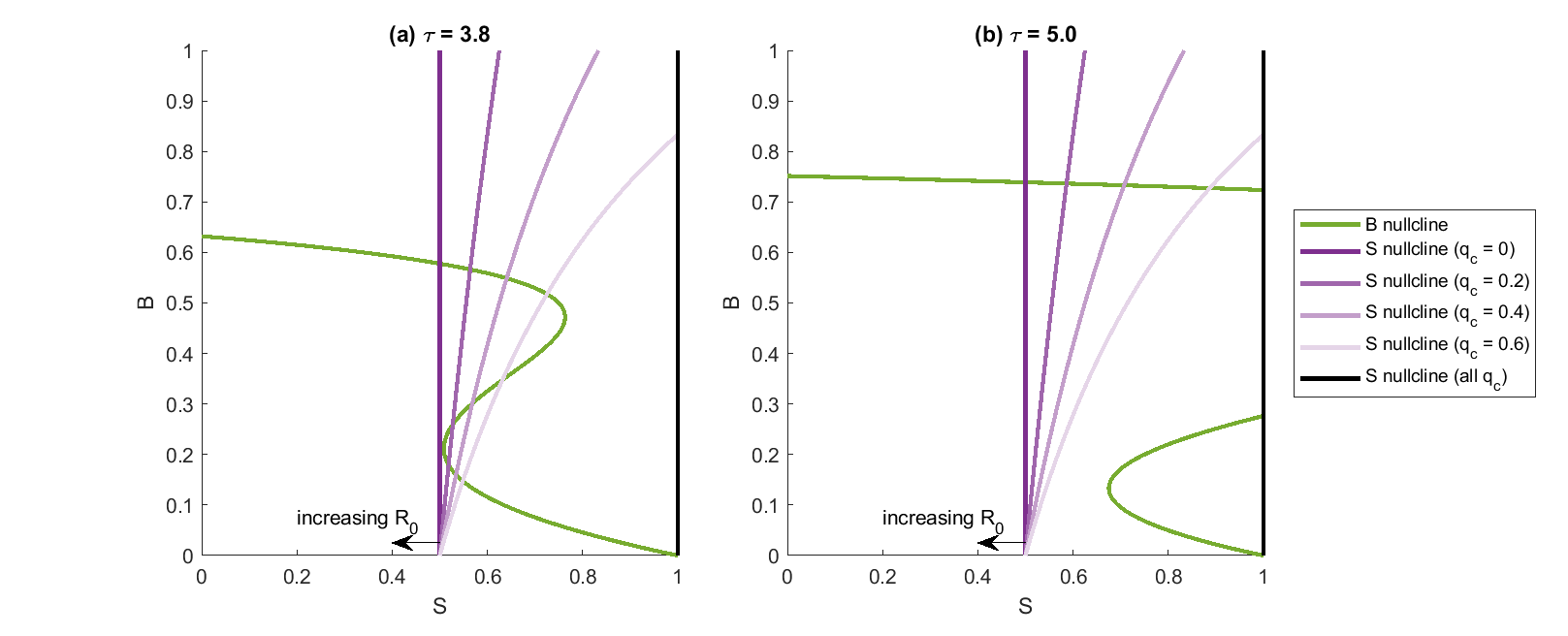}
    \caption{Nullclines of the contagiousness-modulated SIS model in the $(S,B)$ plane for $R_0=2$, various behavioural effect strengths $q_c$ and: (a) $\tau=3.8$; (b) $\tau=5$. Disease-free equilibria occur where the $B$ nullcline (green) and the boundary $S$ nullcline (black) intersect. Endemic equilibria occur where the $B$ nullcline (green) and the interior $S$ nullcline (purple) intersect. The horizontal intercept of the interior $S$ nullcline is at $B=1/R_0=1/2$. Increasing the value of $R_0$ translates the interior $S$ nullcline horizontally to the left; increasing $q_c$ causes it to bend to the right. Saddle node bifurcations of endemic equilibria occur when the $B$ nullcline and interior $S$ nullcline intersect tangentially. Transcritical bifurcations occur when an endemic and a disease-free equilibrium collide, which is when the $B$ nullcline (green), interior $S$ nullcline (purple) and boundary $S$ nullcline (black) all intersect. }
    \label{fig:nullclines}
\end{figure}

Supplementary Figure \ref{fig:nullclines} shows the nullclines (Eqs. \eqref{eq:Snull}--\eqref{eq:Bnull}) of the contagiousness-modulated SIS model in the $(S,B)$ plane. 

\begin{figure}
    \centering
    \includegraphics[trim={0.5cm 0cm 0.5cm 0cm},clip,width=\linewidth]{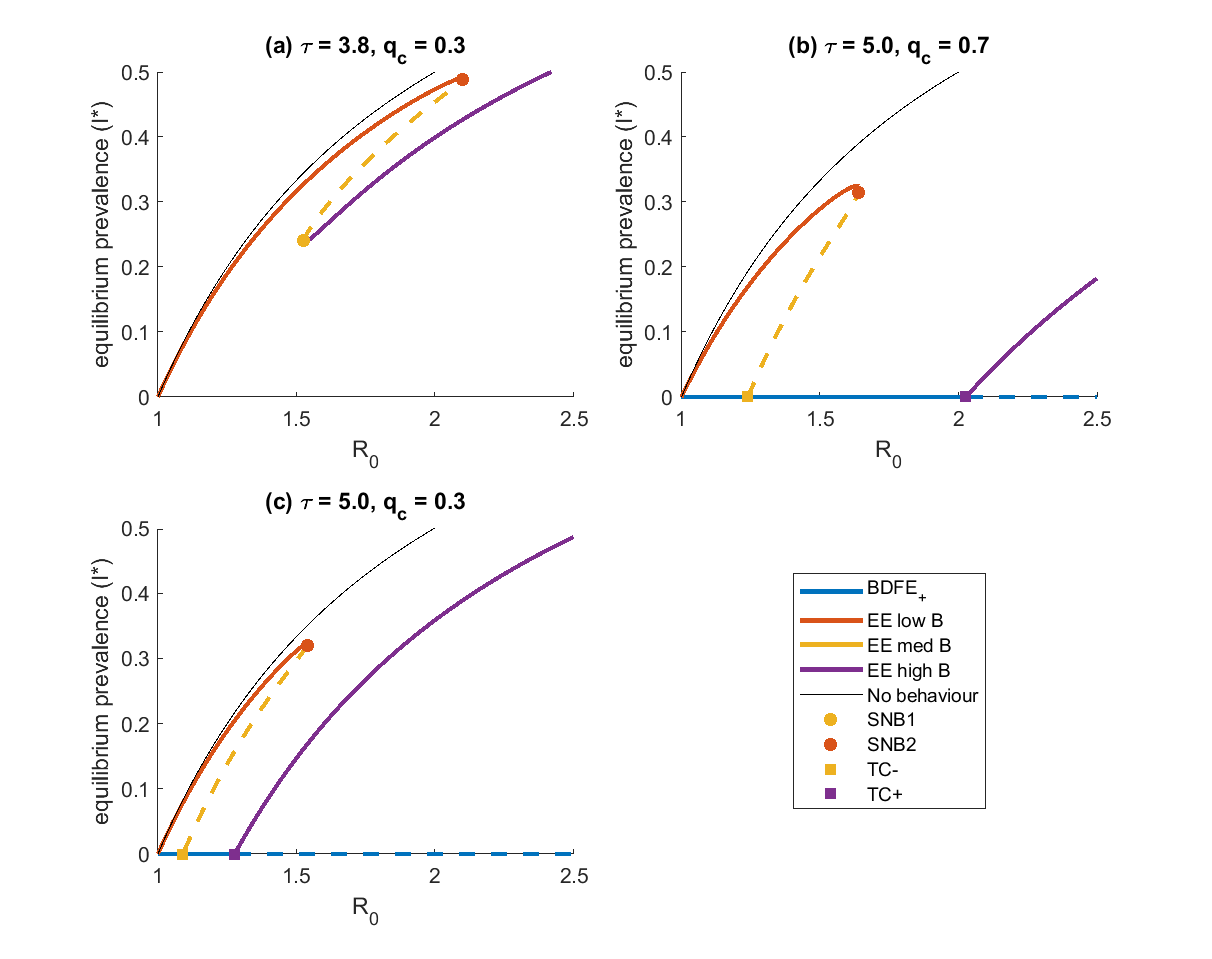}
    \caption{Equilibrium infection prevalence $I^*$ against $R_0$ for the contagiousness-modulated SIS model with $\chi=0.2$ and: (a) $\tau=3.8$ and $q_c=0.3$ (as in Figure 2);  (b) $\tau=5$ and $q_c=0.7$ (as in Figure 3); (c) $\tau=5$ and $q_c=0.3$ (as in Figure 4). Plots show the disease-free equilibrium with high $B$ (BDFE$_+$), the three endemic equilibria with low, medium and high $B$, and for comparison the equilibrium prevalence in the classical SIS model without behavioural dynamics ($I^*=1-1/R_0$). Solid curves show stable equilibria; dashed curves show unstable equilibria. For readability, the NDFE and BDFE$_-$ are not shown (both these equilibria have $I^*=0$). Points show the locations of the saddle node bifurcations SNB1 (when $\tau<4$) and SNB2; and transcritical bifurcations TC$_-$ and TC$_+$ (when $\tau>4$). Note in (b) there is a range of $R_0$ (between SNB2 and TC$_+$) where BDFE$_+$ is the only stable equilibrium (parameter regime 8), meaning that behaviour causes the disease to be eliminated. }
    \label{fig:prevalence_vs_R0}
\end{figure}

Supplementary Figure \ref{fig:prevalence_vs_R0} shows one-parameter ($R_0$)  bifurcation diagrams for the contagiousness-modulated SIS model corresponding to a horizontal cross-section through Figure 1a at $q_c=0.3$, a horizontal cross-section through Figure 1b at $q_c=0.7$, and a horizontal cross-section through Figure 1b at $q_c=0.3$.

\begin{figure}
    \centering
    \includegraphics[trim={0.5cm 0cm 0.5cm 0cm},clip,width=\linewidth]{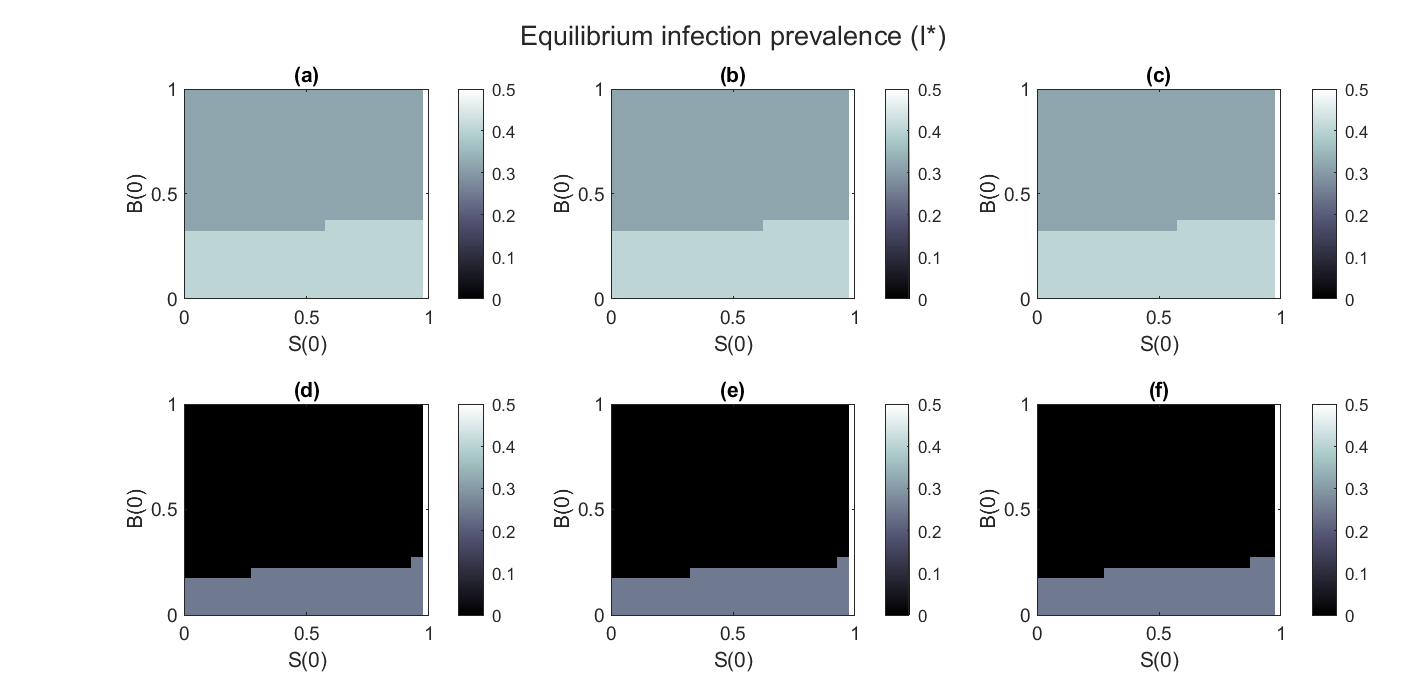}
    \caption{Dependence of equilibrium infection prevalence $I^*$ of the contagiousness-modulated SIS model on initial conditions $S(0)$ and $B(0)$ in bistable cases: (a)--(c) parameter regime 3 ($\tau=3.8$, $q_c=0.3$, $R_0=1.75$); (d)--(f) parameter regime 7 ($\tau=5$, $q_c=0.7$, $R_0=1.4$). Results are shown for three different ways of setting $S_B(0)$: (a,d) model initialised on the invariant two-dimensional manifold by setting $S_B(0)=S(0)B(0)$; (b,e) model initialised with behavers maximally concentrated in the susceptible population by setting $S_B(0)=\mathrm{min}\left(S(0),B(0)\right)$; (c,f) model initialised with behavers maximally concentrated in the infectious population by setting $S_B(0)=\mathrm{max}\left(0,S(0)+B(0)-1\right)$. Other parameter values: $\chi=0.2$, $\alpha=0.25$. }
    \label{fig:IC_dependence}
\end{figure}

Supplementary Figure \ref{fig:IC_dependence} shows the dependence of the equilibrium infection prevalence in the contagiousness-modulated SIS mode on initial conditions, showing that the long-term state is unchanged if the model is initialised away from the two-dimensional manifold defined by $S_B=SB$. 

\begin{figure}
    \centering
    \includegraphics[trim={0.5cm 1.5cm 1cm 0.3cm},clip,width=\linewidth]{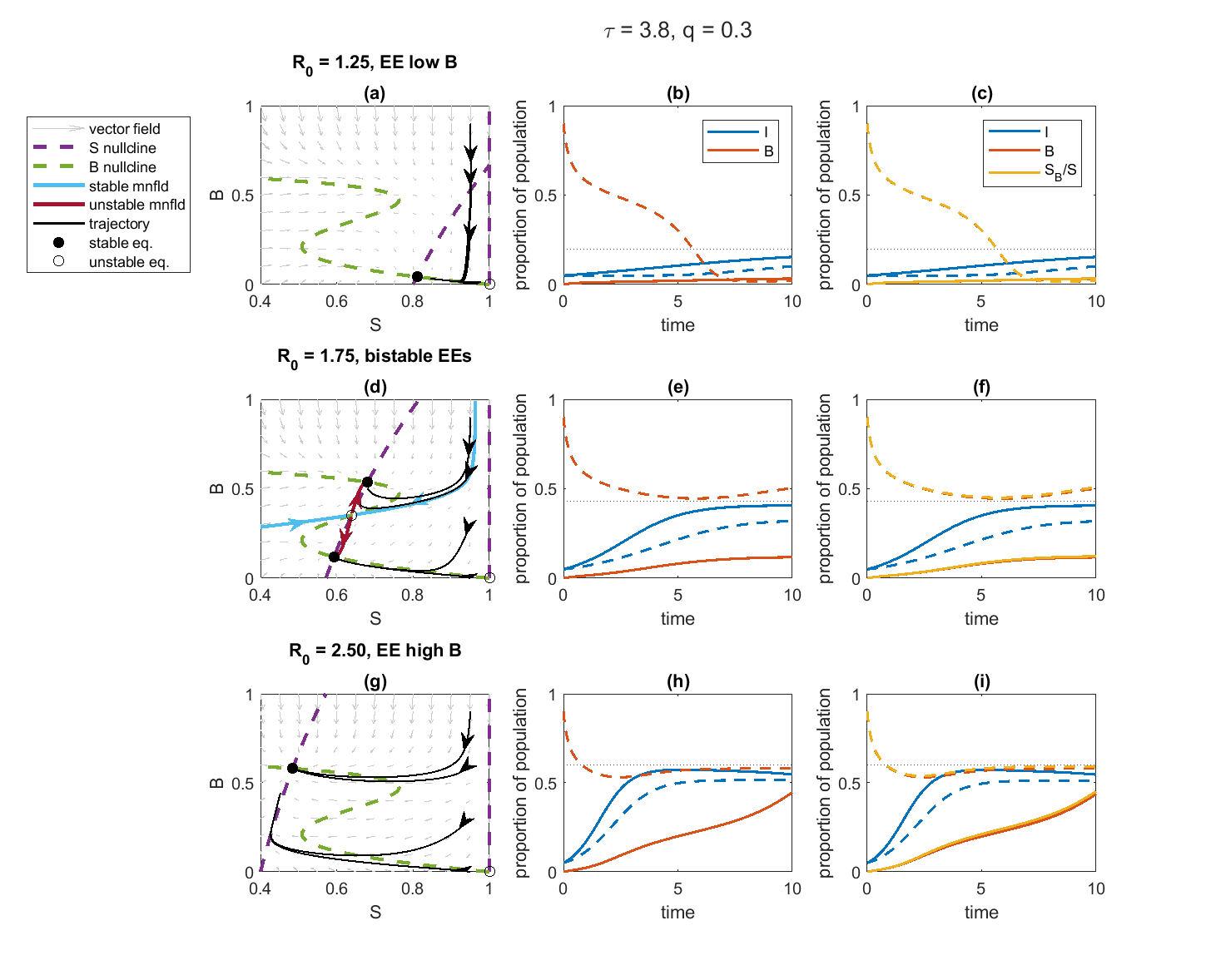}
    \caption{\scriptsize Dynamics in parameter regimes 2 (a,b,c), 3 (d,e,f) and 4 (g,h,i) for: the contagiousness-modulated model with $q_c=0.3$ shown as phase plots in the $(S,B)$ plane (left column) and time series solutions (centre column); and for the susceptibility-modulated model with $q_s=0.3$ shown as time series solutions (right column). All plots have $\tau=3.8$ (so there are no BDFEs), $\chi=0.2$, and $\alpha=2.5$.  Phase plots show the vector field (grey arrows), $S$ nullcline (dashed purple), $B$ nullcline (dashed green), the stable manifold (solid blue) and unstable manifold (solid red) of the saddle endemic equilibrium (when it exists), stable equilibria (filled circles), unstable equilibria (open circles), and trajectories for selected initial conditions (solid black). Time series plots (centre and right columns) show the prevalence of infection (blue) and of behavers (red) over time, for two initial conditions, one with $B(0)=0$ (solid curves) and one with $B(0)=0.9$ (dashed curves). Both initial conditions have $S(0)=0.95$ and $S_B(0)=S(0)B(0)$. Plots for the susceptibility-modulated model (right column) also show the proportion of the susceptible population that is behaving (yellow). In bistable cases, the two initial conditions converge to two different stable equilibria. Otherwise they converge to the unique stable equilibrium.  The dotted horizontal line shows the equilibrium prevalence ($I=1-1/R_0$) for the case where there is no behaviour in the model.}
    \label{fig:supp1}
\end{figure}

Supplementary Figure \ref{fig:supp1} shows SIS model solutions with fast behavioural dynamics ($\alpha=2.5$ instead of $\alpha=0.25$ in Figure \ref{fig:cases1}). Equilibria and their stability are the same as in Figure \ref{fig:cases1}, and the susceptibility-modulated model solutions are very similar to those of the contagiousness-modulated model.

\begin{figure}
    \centering
    \includegraphics[trim={0.5cm 1.5cm 1cm 0.3cm},clip,width=\linewidth]{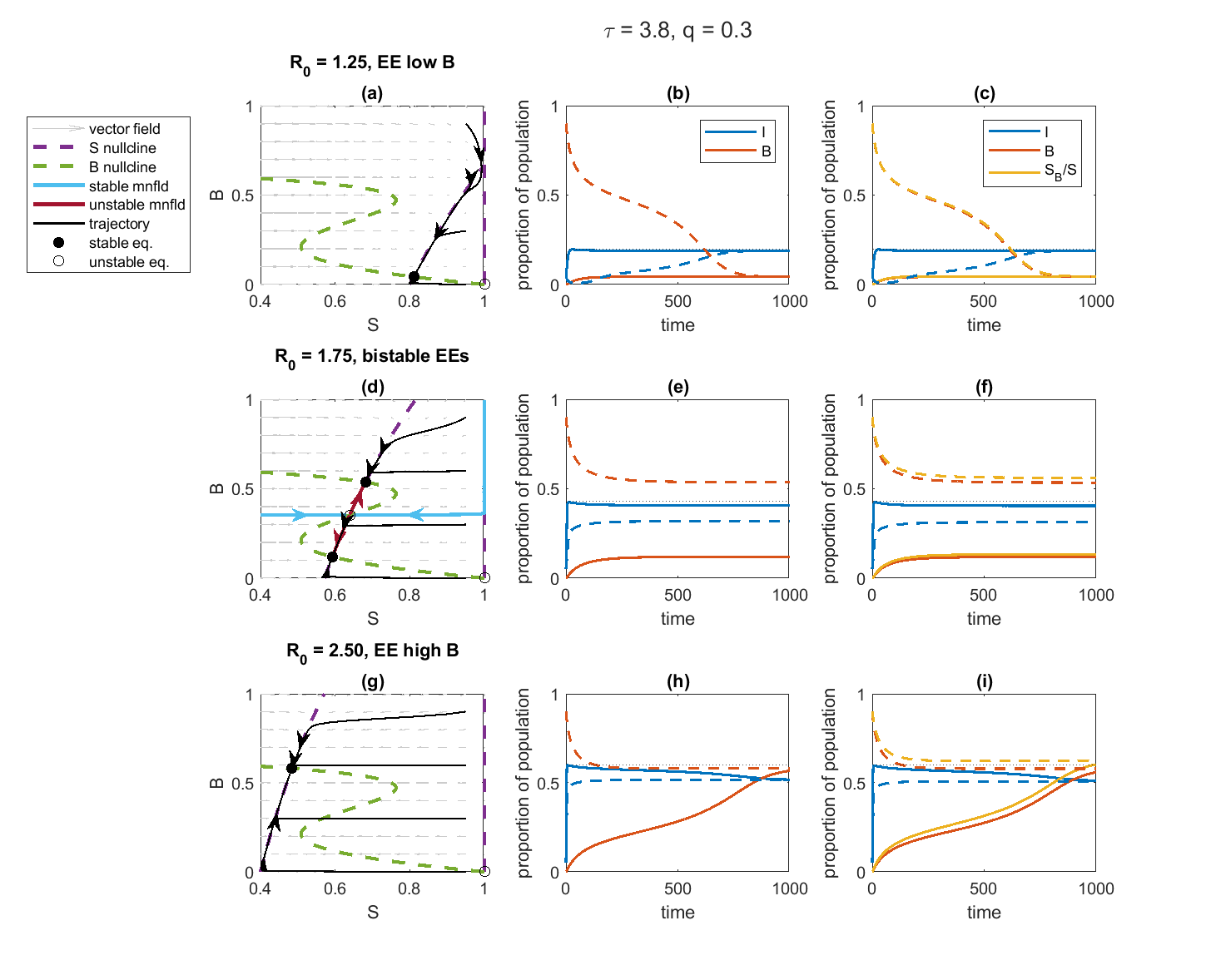}
    \caption{\scriptsize Dynamics in parameter regimes 2 (a,b,c), 3 (d,e,f) and 4 (g,h,i) for: the contagiousness-modulated model with $q_c=0.3$ shown as phase plots in the $(S,B)$ plane (left column) and time series solutions (centre column); and for the susceptibility-modulated model with $q_s=0.3$ shown as time series solutions (right column). All plots have $\tau=3.8$ (so there are no BDFEs), $\chi=0.2$, and $\alpha=0.025$.  Phase plots show the vector field (grey arrows), $S$ nullcline (dashed purple), $B$ nullcline (dashed green), the stable manifold (solid blue) and unstable manifold (solid red) of the saddle endemic equilibrium (when it exists), stable equilibria (filled circles), unstable equilibria (open circles), and trajectories for selected initial conditions (solid black). Time series plots (centre and right columns) show the prevalence of infection (blue) and of behavers (red) over time, for two initial conditions, one with $B(0)=0$ (solid curves) and one with $B(0)=0.9$ (dashed curves). Both initial conditions have $S(0)=0.95$ and $S_B(0)=S(0)B(0)$. Plots for the susceptibility-modulated model (right column) also show the proportion of the susceptible population that is behaving (yellow). In bistable cases, the two initial conditions converge to two different stable equilibria. Otherwise they converge to the unique stable equilibrium. The dotted horizontal line shows the equilibrium prevalence ($I=1-1/R_0$) for the case where there is no behaviour in the model.}
    \label{fig:supp2}
\end{figure}

Supplementary Figure \ref{fig:supp2} shows SIS model solutions with slow behavioural dynamics ($\alpha=0.025$ instead of $\alpha=0.25$ in Figure \ref{fig:cases1}). Equilibria and their stability are the same as in Figure \ref{fig:cases1}, and the susceptibility-modulated model solutions are very similar to those of the contagiousness-modulated model.

\begin{figure}
    \centering
    \includegraphics[trim={0.5cm 1.5cm 1cm 0.3cm},clip,width=\linewidth]{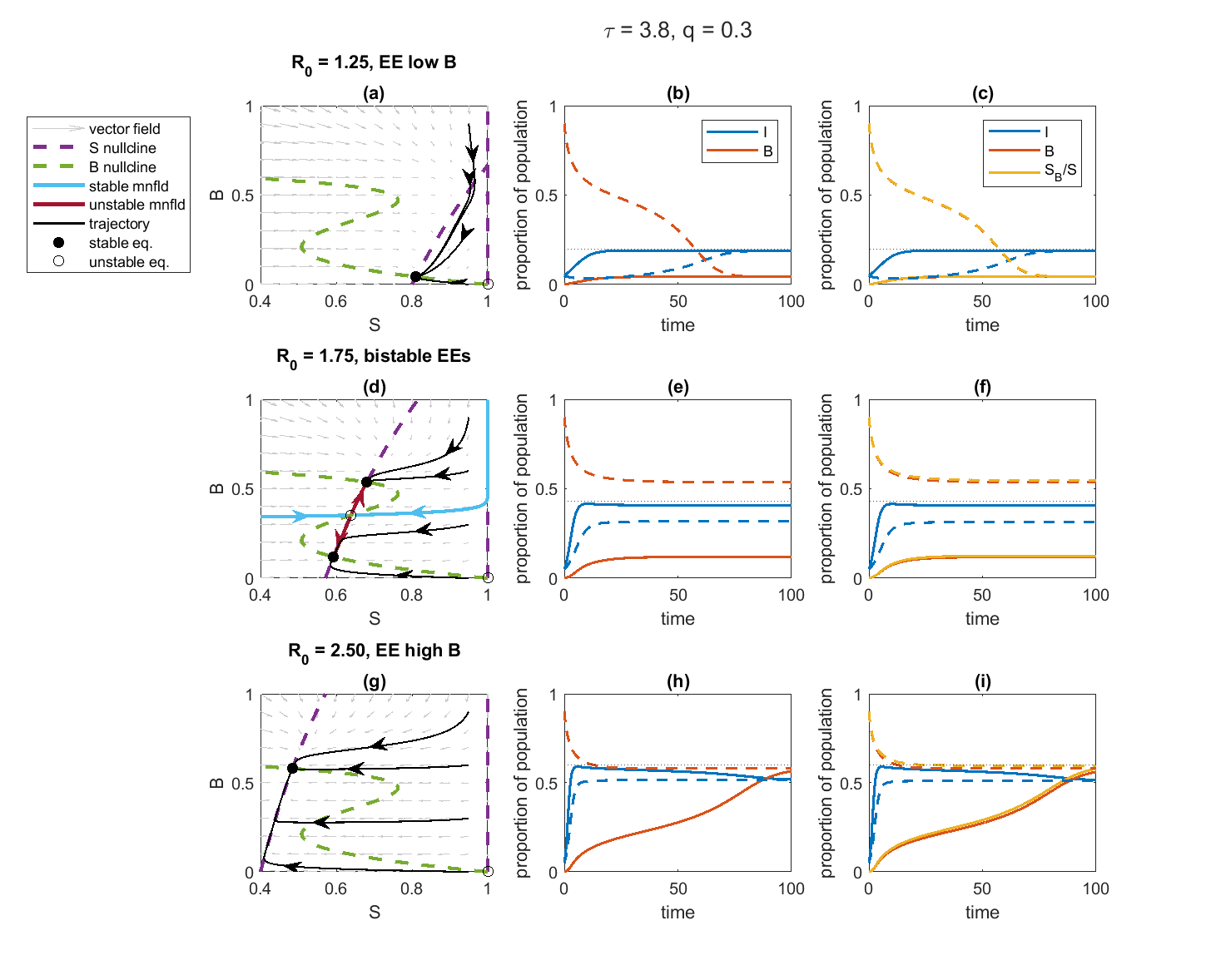}
    \caption{\scriptsize Dynamics in parameter regimes 2 (a,b,c), 3 (d,e,f) and 4 (g,h,i) for: the contagiousness-modulated model with $q_c=0.3$ shown as phase plots in the $(S,B)$ plane (left column) and time series solutions (centre column); and for the general SIS model with $q_c=q_s=0.163$ (so that $1-(1-q_c)(1-q_s)=0.3$) shown as time series solutions (right column). All plots have $\tau=3.8$ (so there are no BDFEs), $\chi=0.2$, and $\alpha=0.25$.  Phase plots show the vector field (grey arrows), $S$ nullcline (dashed purple), $B$ nullcline (dashed green), the stable manifold (solid blue) and unstable manifold (solid red) of the saddle endemic equilibrium (when it exists), stable equilibria (filled circles), unstable equilibria (open circles), and trajectories for selected initial conditions (solid black). Time series plots (centre and right columns) show the prevalence of infection (blue) and of behavers (red) over time, for two initial conditions, one with $B(0)=0$ (solid curves) and one with $B(0)=0.9$ (dashed curves). Both initial conditions have $S(0)=0.95$ and $S_B(0)=S(0)B(0)$. Plots for the general SIS model (right column) also show the proportion of the susceptible population that is behaving (yellow). In bistable cases, the two initial conditions converge to two different stable equilibria. Otherwise they converge to the unique stable equilibrium. The dotted horizontal line shows the equilibrium prevalence ($I=1-1/R_0$) for the case where there is no behaviour in the model.}
    \label{fig:supp3}
\end{figure}

Supplementary Figure \ref{fig:supp3} compares solutions of the contagiousness-modulated SIS model (first and second columns of plots, as in Figure \ref{fig:cases1}) with solutions of the general SIS model where the behaviour affects contagiousness and susceptibility equally ($q_c=q_s=0.163$, so that $1-(1-q_c)(1-q_s)=0.3$). The general SIS model solutions are very similar to those of the contagiousness-modulated model with $q_c=0.3$.

\newpage

\end{document}